\documentclass[twocolumn,showpacs,preprintnumbers,amsmath,amssymb,prd,letterpaper,floatfix,nofootinbib]{revtex4}
\usepackage{graphics}
\usepackage{graphicx}
\usepackage{epsfig}
\usepackage[usenames]{color}
\usepackage{longtable}
\usepackage{hyperref}
\usepackage{latexsym}
\usepackage{amsmath}
\usepackage{amsthm}
\usepackage{amsfonts}
\usepackage{amssymb}
\usepackage{dsfont}
\usepackage{bm}

\newcommand\T{\rule{0pt}{2.6ex}}
\newcommand\B{\rule[-1.2ex]{0pt}{0pt}}

\begin{document}

\title{$D$ and $D_s$ meson spectroscopy}
\author{Daniel Mohler and
R.~M.~Woloshyn\\}
\vspace*{2mm}

\affiliation{
TRIUMF, 4004 Wesbrook Mall Vancouver, BC V6T 2A3, Canada
}

\date{\today}

\begin{abstract}
Results are presented for the low-lying spectrum of $D$ and $D_s$ mesons calculated in lattice QCD using 2+1 flavor Clover-Wilson configurations made available by the PACS-CS collaboration. For the heavy quark, the Fermilab method is employed. The main focus is S- and P-wave states of charmed and charmed-strange mesons, where previous lattice QCD results have been mostly from quenched calculations. In addition to the ground states, some excited states are extracted. To check the method, calculations of the charmonium spectrum are also carried out. For charmonium, the low-lying spectrum agrees favorably with experiment. For heavy-strange and heavy-light systems substantial differences in comparison to experiment values remain for some states.
\end{abstract}


\maketitle

\section{Introduction}

The spectrum of charmed-strange mesons\footnote{Following Particle Data Group usage \cite{0954-3899-37-7A-075021} $D_s$ mesons are referred to as charmed-strange and only $D$ mesons as charmed.} contains a number of well-established states \cite{0954-3899-37-7A-075021}, notably the ``S-wave'' states of quantum numbers $J^P$ $0^-$ (where $J$ is the spin and P is parity)
and $1^-$, the $D_s$ and the $D_s^\star$ and the ``P-wave'' states with quantum numbers $0^+$ ($D_{s0}^\star(2317)$), $1^+$ ($D_{s1}(2460)$ and $D_{s1}(2536)$) and $2^+$ ($D_{s2}^\star(2573)$). In the heavy quark limit, these states would form three mass-degenerate multiplets and are characterized by the total angular momentum $j$ of the light quark. Within many quark models (see for example \cite{Godfrey:1985xj}), the physical states corresponding to the doublet with $j^P=\frac{1}{2}$, the $D_{s0}^\star(2317)$ and $D_{s1}(2460)$ were expected to have considerable width and masses above the $DK$ and $D^\star K$ thresholds. In experiment, both states are below threshold and narrow. 

In addition there are a number of $D_s$ states which have been observed more recently. These are the $D_{s1}^\star (2700)$ (observed by both BaBar \cite{Aubert:2006mh,Aubert:2009di} and Belle \cite{Brodzicka:2007aa}), the $D_{sJ}^\star(2860)$ (observed by BaBar \cite{Aubert:2006mh,Aubert:2009di}), the $D_{sJ}(3040)$) (observed by BaBar \cite{Aubert:2009di}) and an unconfirmed state previously observed by SELEX \cite{Evdokimov:2004iy}, the $D_{sJ}(2632)$. While the $D_{s1}^\star (2700)$ is commonly believed to have quantum numbers $J^P=1^-$, there are several possibilities for the other states which are not ruled out by experiment. The $D_{sJ}^\star(2860)$ has natural parity and is most often identified with a $3^-$ state \cite{Colangelo:2010te,Ebert:2009ua,Chen:2009zt}, while some still argue the possibility of a $0^+$ identification \cite{vanBeveren:2009jq}. The $D_{sJ}(3040)$ has unnatural parity and is commonly interpreted as either a $1^+$ \cite{Colangelo:2010te,Ebert:2009ua,Sun:2009tg,Chen:2009zt} or a $2^-$ state \cite{Colangelo:2010te}.

Until recently, the number of observed charmed meson states was more limited. Again the ``S-wave'' states of quantum numbers $J^P$ $0^-$ ($D$)
and $1^-$ ($D^\star$) as well as the ``P-wave'' states with quantum numbers $0^+$ ($D_0^\star(2400)$),  $1^+$ ($D_1(2420)$ and $D_1(2430)$) and $2^+$ ($D_2^\star(2460)$) are well-established \cite{0954-3899-37-7A-075021}. In addition, the Particle Data Group \cite{0954-3899-37-7A-075021} lists the $D^\star(2640)$ seen in Z-decays \cite{Abreu:1998vk} which lacks confirmation. More recently BaBar observed several new charmed mesons \cite{delAmoSanchez:2010vq} whose quantum numbers are a matter of active discussion \cite{Sun:2010pg,Li:2010vx,Wang:2010ydc,Zhong:2010vq,Chen:2011rr}.

Lattice QCD (LQCD) provides the possibility to elucidate the spectrum without resorting to model assumptions. To reach this goal, several systematic sources of uncertainty have to be controlled. Of particular importance is that full dynamical simulations are necessary with up and down quarks light enough so that an extrapolation to the physical point is not needed or can be done in a meaningful way. Also, there should be an extrapolation to the continuum limit. Such calculations have been achieved for the light-quark ground state meson and baryon spectrum (see, for example, \cite{Durr:2008zz}). 

Dealing with heavy charm or bottom quarks on the lattice introduces complications of its own, and so far a similar precision for the spectrum of charmed and charmed-strange states has not been attained. Most previous lattice calculations have been within the quenched approximation or at unphysically large pion masses. Advances in lattice simulations enable us to revisit this problem, removing one deficiency of previous calculations by using dynamical gauge ensembles where pion masses are close to the physical pion mass. 

Among the low-lying $D$ and $D_s$ mesons accessible to current lattice calculations, the $D_{s0}^\star(2317)$ and $D_{s1}(2460)$, whose properties can not be explained within most quark models, are of particular interest. Possible explanations for their unanticipated mass and width have been reviewed in \cite{Swanson:2006st,Zhu:2007wz}. In early LQCD calculations, the ground states in the $0^+$ and $1^+$ channels were often found to be quite a bit heavier than the experimental resonances (for a short summary of previous results please refer to Section \ref{hl_results}). This has further sparked speculations about the nature of these states and provides motivation for this study.

The next section provides some details of our calculational setup. There is also a short description of the procedure used to tune the charm quark mass. To test our setup, the spectrum of charmonium states below multiparticle threshold was calculated and is discussed in Section \ref{charmonium}. Section \ref{hl_results} presents our results and discussion of the charmed and charmed-strange mesons. Finally, in the last section, a summary of our findings is presented. Also included is a short description of the role of scattering thresholds for the calculation. Some preliminary results have already been presented in \cite{Mohler:2010nj}.

\section{\label{calc}Calculational setup}

A subset of the dynamical 2+1 flavor configurations generated with the Clover-Wilson fermion action by the PACS-CS collaboration \cite{Aoki:2008sm}, spanning sea-quark pion masses from 702 MeV down to 156 MeV, are used in this work\footnote{After work on this project started dynamical simulations with even smaller pion masses, for the first time at \cite{Aoki:2009ix} and around \cite{Durr:2010vn,Durr:2010aw} the physical point, were reported.}. The number of lattice points is $32^3\times64$ for all ensembles, and using the single lattice spacing of 0.0907(13) fm, as determined in \cite{Aoki:2008sm} this corresponds to a box size of roughly 2.9fm in spatial direction and 5.8fm in time direction. Table \ref{paratable} shows the parameters for the runs. With exception of the ensemble \#5 we use $\sim200$ configurations on each ensemble.

\begin{table}[bht]
\begin{center}
\begin{ruledtabular}
\begin{tabular}{|c|c|c|c|c|}
 Ensemble & $c_{sw}^{(h)}$ & $\kappa_{u/d}$ &  $\kappa_s$  & \#configs $D/D_s$\\
\hline
 1 & 1.52617 & 0.13700 & 0.13640 & 200/200\\
 \hline
2 & 1.52493 & 0.13727 & 0.13640 & -/200\\
\hline
 3 & 1.52381 & 0.13754 & 0.13640 & 200/200\\
\hline
 4 & 1.52327 & 0.13754 & 0.13660 & -/200\\
\hline
 5 & 1.52326 & 0.13770 & 0.13640 & 200/348\\
\hline
 6 & 1.52264 & 0.13781 & 0.13640 & 198/198\\
\end{tabular}
\end{ruledtabular}
\end{center}
\caption{\label{paratable}Run parameters for the PACS-CS lattices \cite{Aoki:2008sm}. The number of lattice points is $32^3\times64$ for all ensembles. All gauge configurations have been generated with the inverse gauge coupling $\beta=1.90$ and the light quark clover coefficient $c_{sw}^{(l)}=1.715$. The number of configurations indicated in the table corresponds to the number used rather than the total number available. The quantity $c_{sw}^{(h)}$ denotes the heavy quark clover coefficient. Please refer to Section \ref{tuning} for the determination of $c_{sw}^{(h)}$.}
\end{table}

\subsection{\label{sources}Source construction}

The low-lying spectrum is extracted using the variational method \cite{Luscher:1990ck,Michael:1985ne}. For each combination of quantum numbers $J^P$ (where $J$ is the Spin and $P$ is parity) a matrix $C(t)_{ij}$ of interpolators is constructed
\begin{align}
C(t)_{ij}&=\sum_n\mathrm{e}^{-tE_n}\left <0|O_i|n\right>\left<n|O_j^\dagger|0 \right> .
\end{align}
Subsequently, the the generalized eigenvalue problem is solved for each time slice
\begin{align}
C(t)\vec{\psi}^{(k)}&=\lambda^{(k)}(t)C(t_0)\vec{\psi}^{(k)} ,\\
\lambda^{(k)}(t)&\propto\mathrm{e}^{-tE_k}\left(1+\mathcal{O}\left(\mathrm{e}^{-t\Delta E_k}\right)\right).\nonumber
\end{align}
At large time separation only a single mass contributes to each eigenvalue. $\Delta E_k$ is in general given by the energy difference between the energy level in consideration and the neighboring level. For more details the reader is referred to the discussion in \cite{Blossier:2009kd}. For the determination of the charm quark hopping parameter we will use projections to the lowest few lattice momenta, while masses are obtained from a projection to vanishing spatial momentum. In addition to the eigenvalues, the eigenvectors  provide useful information and can serve as a fingerprint for a given state.

The basis is constructed from two types of sources. The first type are simple Jacobi-smeared \cite{Gusken:1989ad,Best:1997qp} Gaussian-shape sources $q_s\equiv (S \,q)_x$
\begin{align}
S&=M \; S_0\quad\mbox{with}\quad M=\sum_{n=0}^N\kappa^nH^n ,\nonumber\\
H(\vec{n},\vec{m}\,)&=\sum_{j = 1}^3
\left(U_j\left(\vec{n},0\right) \delta\left(\vec{n} + \hat{j}, \vec{m}\right) \right.\\
&\quad\left.+ U_j\left(\vec{n}-\hat{j\,},0\right)^\dagger 
\delta\left(\vec{n} - \hat{j}, \vec{m}\right) 
\right) . \nonumber
\end{align}
Here, $H$ is the spatial hopping term containing the link variables $U_j(\vec{n},t)$ and the smearing has two parameters $\kappa$ and $N$. In addition, we also use derivative sources $W_{d_i}$
\begin{align}
W_{d_i}&=D_i\,S\
,\\
D_i(\vec{x},\vec{y})&=U_i(\vec{x},0)\delta(\vec{x}+\hat{i},\vec{y})-U_i(\vec{x}-\hat{i},0)^\dagger\delta(\vec{x}-\hat{i},\vec{y})\ .\nonumber
\end{align}
The parameters for the Gaussian-shape sources are  $\kappa=0.22$ and $N=25$. For the derivative sources the number of iterations is slightly larger $N=30$ while using the same $\kappa$. The full list of interpolating fields used in the simulation can be found in Appendix \ref{source_tables}. Similar source constructions have been used for both quarkonium \cite{Liao:2002rj,Dudek:2007wv} and light meson \cite{Lacock:1996vy,Gattringer:2008be,Dudek:2009qf} spectroscopy. To minimize the calculation of expensive propagators, we use both Gaussian and derivative sources for the charm quark but only Gaussian sources for the strange and light quarks. This has an additional implication: The interpolators with derivative sources are constructed from a derivative which is not symmetrically applied to both quarks and these interpolators do not have good charge conjugation quantum number. This is relevant in the case of charmonium, where the same basis is used. For further comments regarding charge conjugation, please refer to Section \ref{charmonium}.

\subsection{Quark propagators}

For the construction described in the previous section we need a single full propagator for the strange (light) quark and 4 full propagators (one Gaussian and three derivatives) for the charm quark. These propagators are calculated for 8 source time slices on each gauge configuration. The source locations are chosen randomly within the time slice. For the calculation of the light and strange quark propagators the dfl\_sap\_gcr inverter from L\"uscher's DD-HMC package \cite{Luscher:2007se,Luscher:2007es} is employed. For the charm quark propagators the corresponding inverter without deflation is used. The large number of sources needed for this work makes the use of a deflation inverter extremely efficient, particularly for light quarks on Ensemble 6, with a pion mass of 156MeV.

\subsection{Fitting methodology}

This section presents the methodology for fits to both single correlators as well as eigenvalues from the variational method. All of these fits take into account autocorrelation in Euclidean time $t$. The estimate of the covariance matrix Cov$(t,t^\prime)$ employs a single-elimination jackknife
\begin{align}
\mathrm{Cov}(t,t^\prime)&=\frac{N-1}{N}\sum_{i=1}^N\left(\bar{\lambda}^{(i)}(t)-\bar{\lambda}(t)\right)\left(\bar{\lambda}^{(i)}(t^\prime)-\bar{\lambda}(t^\prime)\right) ,
\end{align} 
where N denotes the number of configurations and the bars denote averages. To obtain a stable estimate, the window for the inversion of the covariance matrix is restricted to the chosen fit range. The inverse covariance matrix is estimated once on the ensemble average and used for each jackknife block. This method has been dubbed ``jackknife reuse'' in \cite{Toussaint:2008ke}. To determine the window in Euclidean time for performing fits we calculate effective masses
\begin{align}
aM_{\mathrm{eff}}&=\mathrm{ln}\left(\frac{\lambda^{(i)}(t)}{\lambda^{(i)}(t+1)}\right)
\end{align}
and eigenvector components of the regular eigenvalue problem
\begin{align}
C(t_0)^{-\frac{1}{2}}C(t)C(t_0)^{-\frac{1}{2}}\vec{\psi}^{(k)\,\prime}&=\lambda^{(k)}(t)\vec{\psi}^{(k)\,\prime} ,
\end{align}
and choose as our fit range the interval where both the effective masses and the eigenvector components are constant within their statistical uncertainties. Then two parameter fits are carried out within this window. As a cross check, 4 parameter fits with two exponentials and a larger fit range are performed. In some cases, the 4 parameter fits are unstable. Where stable, the 4 parameter fits lead to results that are consistent within errors with the strategy described above. The dependence of the results on the time slice $t_0$ was investigated and, where necessary, the reference time $t_0$ was increased until the results were qualitatively unaffected by a further increase. Appendix \ref{tables_b} collects various fit results, fit intervals and the associated $\chi^2/\mathrm{d.o.f.}$.

Due to the method employed for the heavy quark, the energy levels extracted should not be identified with the physical energy levels directly. Instead, as final results we always quote energy differences with respect to the spin-averaged 1S states, given by
\begin{align}
M_{\overline{D_s}}&=(M_{D_s}+3M_{D_s^\star})/4 ,\nonumber\\
M_{\overline{D}}&=(M_{D}+3M_{D^\star})/4 ,\\
M_{\overline{\bar{c}c}}&=(M_{\eta_c}+3M_{J/\Psi})/4 ,\nonumber
\end{align} 
for the case of heavy-strange mesons, heavy-light mesons and charmonium respectively. As the measurements are on the same gauge ensemble and therefore correlated, taking into account this correlation leads to reduced statistical errors for the mass differences.

\subsection{\label{tuning}Heavy quark action and quark mass tuning}

To determine the mass parameter for the heavy charm quark we use the \emph{Fermilab method} \cite{ElKhadra:1996mp} in the form employed by the Fermilab Lattice and MILC collaborations \cite{Bernard:2010fr} for their efforts involving quarkonium \cite{Burch:2009az}. This section describes this method in brief and points out where our method differs from \cite{Bernard:2010fr}. In addition to this simplest prescription, an improved Fermilab action has been constructed \cite{Oktay:2008ex} and preliminary results are encouraging \cite{Detar:2010aq}. Within this approach, the charm quark hopping parameter $\kappa_c$ is tuned to the value where the spin-averaged \emph{kinetic mass} $(M_{D_s}+3M_{D_s^*})/4$ assumes its physical value. In this simplest formulation the heavy quark hopping parameter $c_E=c_B=c_{sw}^{(h)}$ is set to its tadpole improved \cite{Lepage:1992xa} value $\frac{1}{u_0^3}$, where $u_0$ is the average link. For simplicity, $u_0$ is estimated to be the fourth root of the average plaquette, unlike in \cite{Bernard:2010fr}, where it is given by the Landau link. The lattice dispersion relation takes the general form \cite{Bernard:2010fr}
\begin{align}
E(p)&=M_1+\frac{\mathbf{p}^2}{2M_2}-\frac{a^3W_4}{6}\sum_ip_i^4-\frac{(\mathbf{p}^2)^2}{8M_4^3}+ \dots ,
\label{disp}
\end{align}
where the momentum $\mathbf{p}=\frac{2\pi}{L}\mathbf{n}$ and $L$ is the spatial extent of the lattice. 
To determine the kinetic mass $M_2$ the six lowest lattice momenta are used and averaged over all possible vectors $\mathbf{n}$. These are denoted symbolically as 000, 100, 110, 111, 200 and 210. Larger momenta are very noisy and do not help to constrain the fits further. As there are not enough points to constrain a fit to the four-parameter form of Equation \ref{disp}, two simplified methods are considered:
\begin{itemize}
\item [1.] neglect the term with coefficient $W_4$ and fit $M_1$, $M_2$ and $M_4$.
\item [2.] fit $E^2(p)$ and neglect the $(p^2)^2$ term arising from the mismatch of $M_1$, $M_2$ and $M_4$
\begin{align}
E^2(p)&\approx M_1^2+\frac{M_1}{M_2}\mathbf{p}^2-\frac{M_1a^3W_4}{3}\sum_i(p_i)^4 .
\end{align}
\end{itemize}

\begin{figure}[tbh]
\includegraphics[width=85mm,clip]{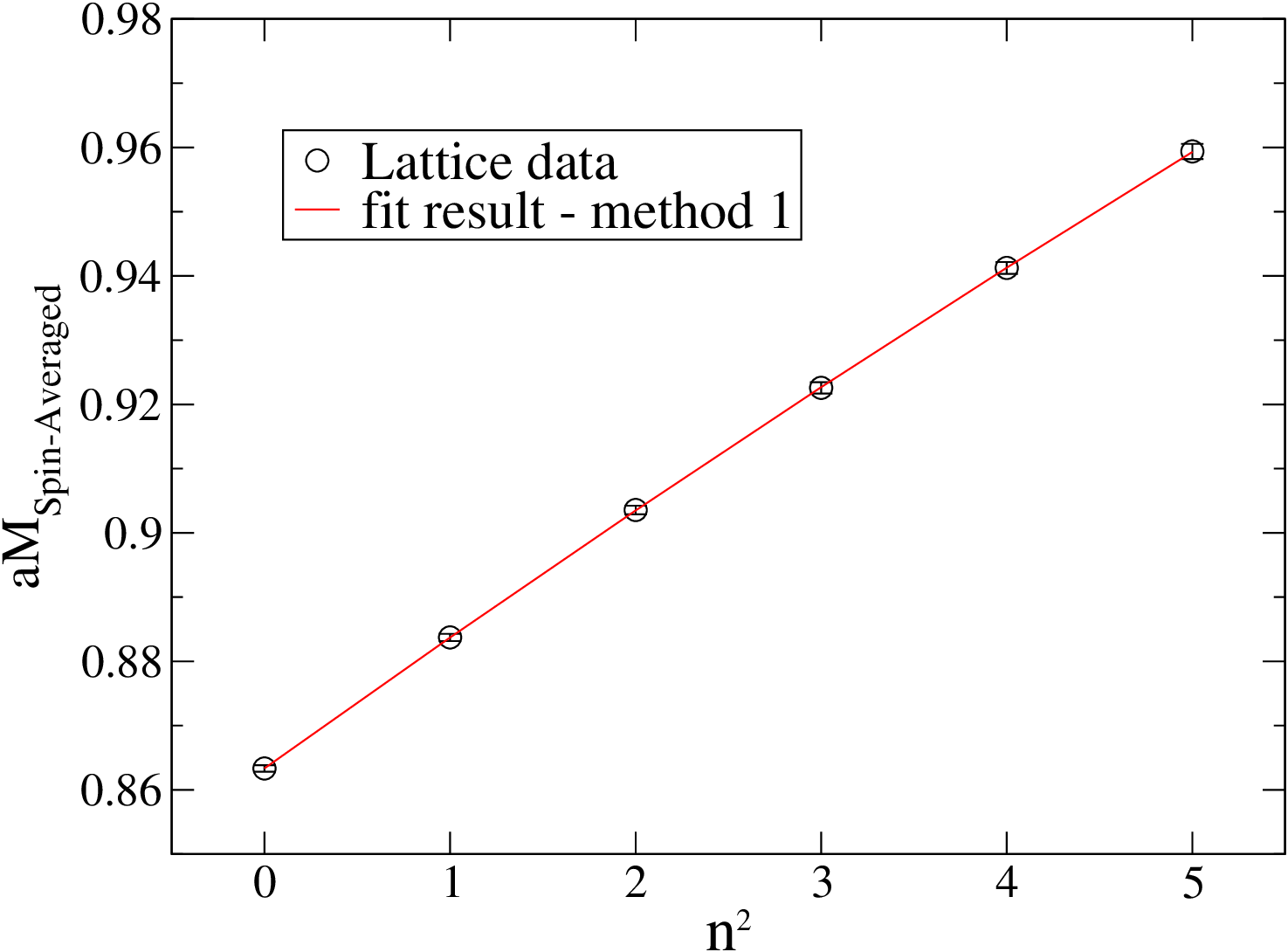}
\caption{Example fit using method 1 with $\kappa_c=0.128$} 
\label{example_tuning}
\end{figure}

For both methods, two values of the charm quark hopping parameter close to the physical value are used and the results are interpolated linearly. To determine the energy levels, a $2\times 2$ matrix of interpolating fields corresponding to interpolators A and B in Appendix \ref{source_tables} is used in both the $J^P=0^-$ and $1^-$ sectors. We choose $t_0=3$ for these fits. Table \ref{firstmethod} shows the results for the spin-averaged ground state from the first method. For illustration an example fit from this method  with $\kappa_c=0.128$ is shown in Figure \ref{example_tuning}.

\begin{table}[bht]
\begin{center}
\begin{ruledtabular}
\begin{tabular}{|c|c|c|}
 \T\B & $\kappa_c=0.128$ & $\kappa_c=0.127$\\
\hline
\T\B $M_1$ & 0.8633(5) & 0.8931(5)\\
\hline
\T\B $M_2$ & 0.9337(73) & 0.9716(76)\\
\hline
\T\B $M_4$ & 0.8638(274) & 0.8855(284)\\
\hline
\T\B $\frac{M_2}{M_1}$ & 1.0815(86) & 1.0878(88)\\
\hline
\hline
\T\B $M_2 [GeV]$ & 2.0315(158)(291) & 2.1137(166)(303)\\
\end{tabular}
\end{ruledtabular}
\end{center}
\caption{\label{firstmethod}Fit parameters obtained with our first method for two different values of $\kappa_c$. The values in the last row are in GeV, while all other values are in lattice units. The first error on the kinetic mass $M_2$ is statistical while the second error is from the scale setting. The results for $M_4$ are not used in our setup.}
\end{table}

The three parameter fits are stable and both the rest mass $M_1$ and the kinetic mass $M_2$ are well-determined. The mass $M_4$ is not used in our analysis but it is encouraging that the result is consistent with the expectation $M_1\sim M_2\sim M_4$. The error on the kinetic mass in physical units has two contributions and the last row of Table \ref{firstmethod} lists both the statistical uncertainty and the uncertainty from setting the lattice scale. This second error from the conversion to physical units is dominant for method 1.

\begin{table}[bht]
\begin{center}
\begin{ruledtabular}
\begin{tabular}{|c|c|c|}
 \T\B & $\kappa_c=0.128$ & $\kappa_c=0.127$\\
\hline
\T\B $M_1$ & 0.8634(5) & 0.8932(5)\\
\hline
\T\B $\frac{M_2}{M_1}$ & 1.0889(116) & 1.0955(118)\\
\hline
\hline
\T\B $M_2 [GeV]$ & 2.0454(215)(293) & 2.1293(227)(305) \\
\end{tabular}
\end{ruledtabular}
\end{center}
\caption{\label{secondmethod}Fit parameters obtained with our second method for two different values of $\kappa_c$. The values in the last row are in GeV, while all other values are in lattice units. The first error on the kinetic mass $M_2$ is statistical while the second error is from the scale setting.}
\end{table}

\begin{figure}[tbh]
\includegraphics[width=85mm,clip]{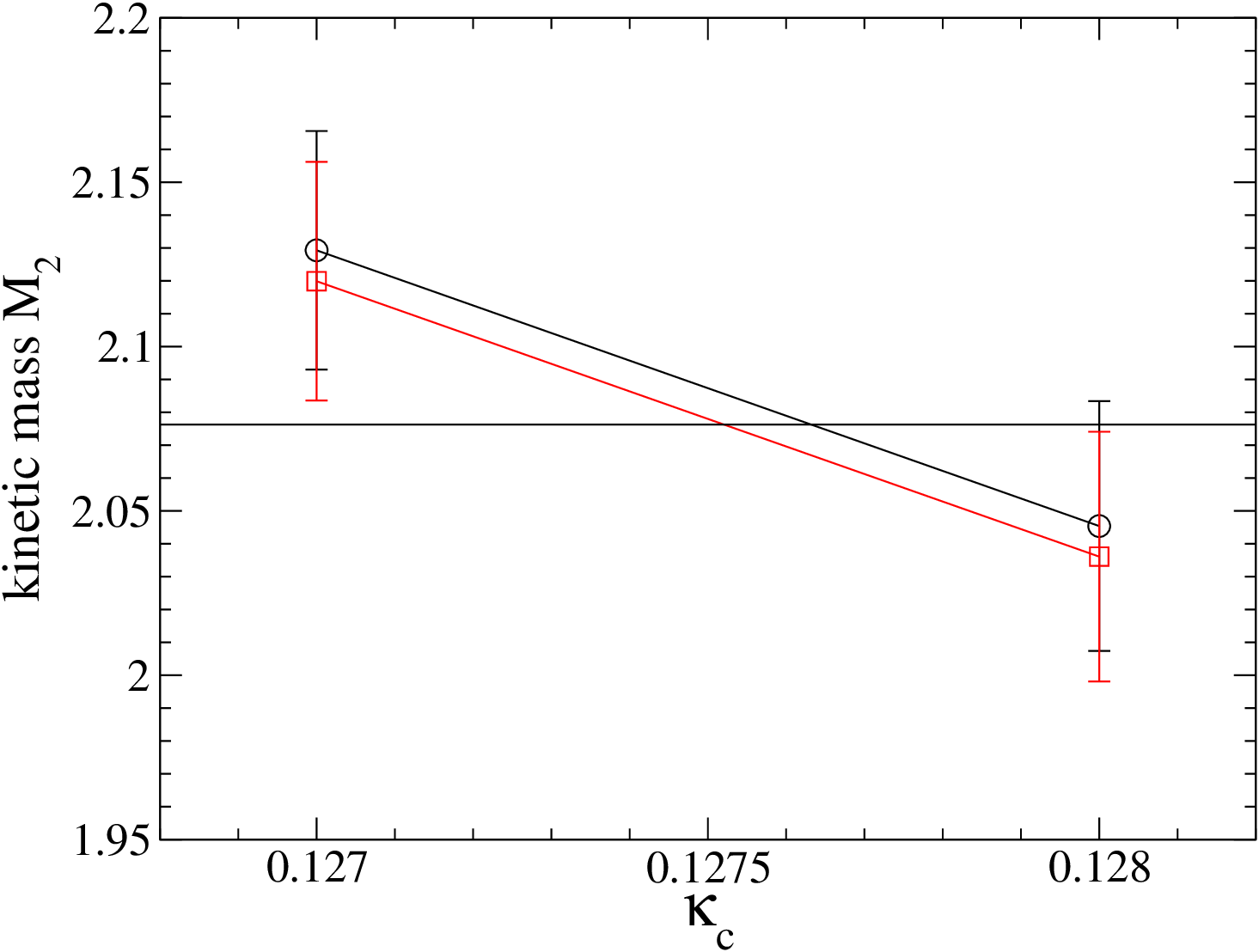}
\caption{Linear interpolation to determine the charm quark hopping parameter $\kappa_c$. The horizontal line corresponds to the physical value for $(M_{D_s}+3M_{D_s^\star})/4$. The red curve (squares) is a shift taking into account the slightly unphysical strange quark mass.} 
\label{kappa_interpolation}
\end{figure}

The results from method two can be found in Table \ref{secondmethod}. This second approach leads to slightly larger values for the kinetic mass $M_2$. The results of the two methods are however consistent within the statistical uncertainties, which are slightly larger for the second method. With both methods $M_2$ is determined sufficiently well, in the sense that the error due to the lattice scale setting is dominant. The calculation of the heavy-light and charmonium spectra in Sections \ref{charmonium} and \ref{hl_results} uses the estimate of $\kappa_c$ from fits with the second method. For this, a plot of the results for the spin-averaged ground state is shown in Figure \ref{kappa_interpolation}. To account for the slightly unphysical strange quark mass used in the simulation we also calculated the spin-averaged rest mass with a partially quenched strange quark mass corresponding to the estimate of the physical strange quark mass from \cite{Aoki:2008sm}. The red curve (squares) in Figure \ref{kappa_interpolation} results from shifting our original results to account for this difference. One  can then proceed to read off the value $\kappa_c=0.12752$ with which the simulations to determine the low-lying spectrum are performed.

Having tuned the charm quark hopping parameter using $D_s$ mesons, a stringent test of our setup can be performed by calculating the low-lying charmonium spectrum, where there are several well-established states below multiparticle-threshold.

\section{\label{charmonium}Charmonium results}

In recent years many new charmonium resonances have been observed, primarily at the B-factories. While the agreement between the well-established low-lying states and quark model calculations (see for example \cite{Godfrey:1985xj}) is quite good, many of the newly observed X, Y and Z states do not fit within a simple quark model picture. For a recent review regarding these charmonium-like states please refer to \cite{Godfrey:2008nc}. For an emphasis on recent results see \cite{Godfrey:2009qe,Biassoni:2010ew,Lange:2010dp}. The charmonium states considered in this section are all well established and considered to be predominantly regular quark-antiquark states with masses well below the $DD$ and $DD^\star$ thresholds. This makes the calculation of the low-lying charmonium spectrum a good test of the heavy-quark methodology. Where appropriate, we somewhat sloppily refer to those states with their spectroscopic nomenclature and call them  S- and P-wave states respectively.

 As mentioned in Section \ref{sources} some of our interpolators, which were chosen with heavy-light mesons in mind, mix different charge conjugations. In the case of the $\eta_c$ ($J^{PC}=0^{-+}$), the $J/\Psi$ ($J^{PC}=1^{--}$) and the $\chi_{c0}$ channels ($J^{PC}=0^{++}$) the corresponding admixtures would come from exotic states, which are expected to play a role higher in the spectrum but not in the vicinity of the low-lying states extracted in this section. In the case of the $1^+$ channel(s) the basis is restricted to those interpolators with good charge conjugation and for the interpolators containing the $\chi_{c2}$ we just extract the ground state, which again should be safe from possible contamination.

\begin{figure}[tbp]
\includegraphics[width=85mm,clip]{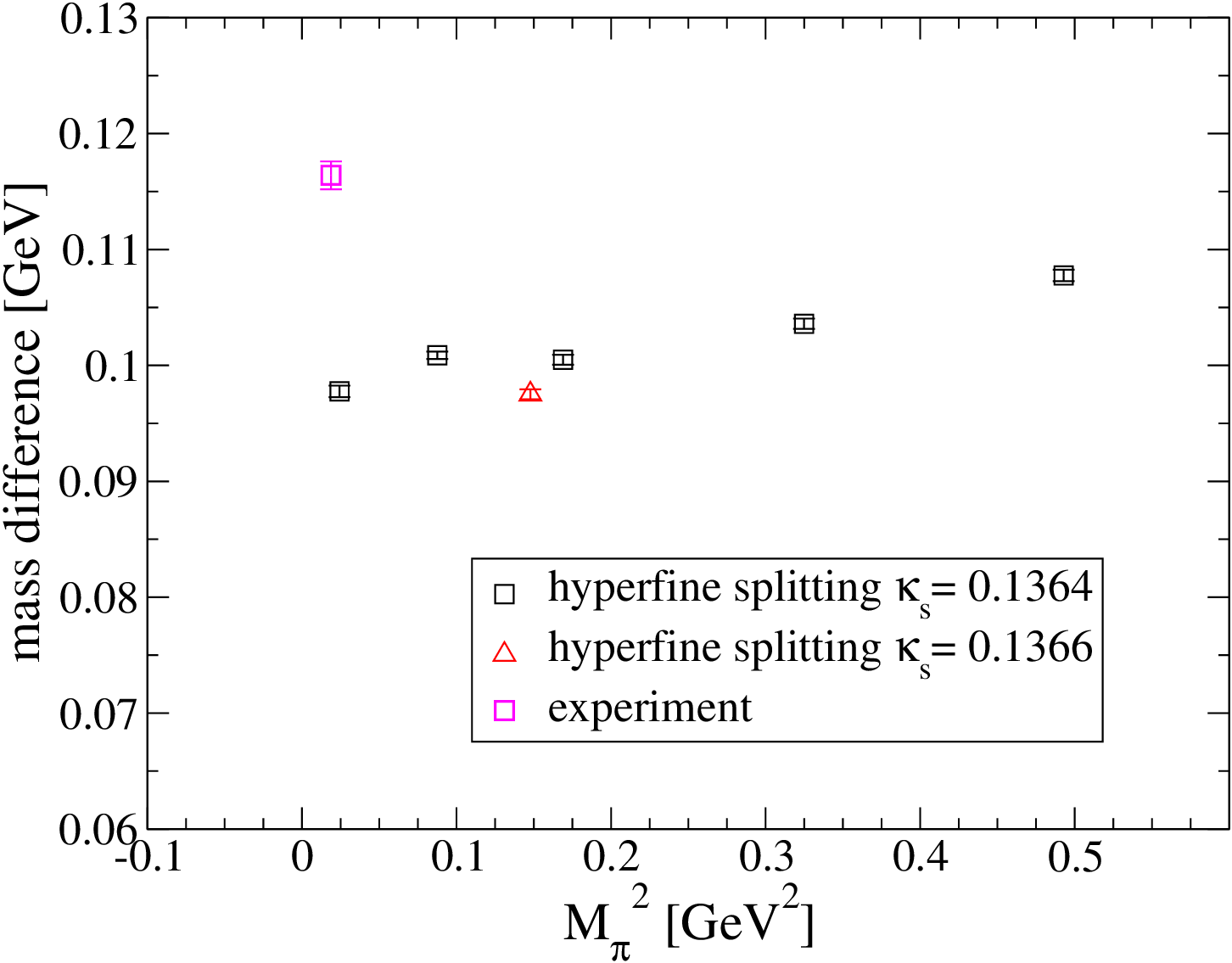}
\caption{1S hyperfine-splitting for charmonium compared to the physical splitting. The error is statistical only and the leading discrepancy is caused by the lack of a continuum extrapolation. For further discussion, please refer to the text.} 
\label{hyperfine_charmonium}
\end{figure}

Figure \ref{hyperfine_charmonium} shows the chiral behavior of the charmonium hyperfine-splitting $M_{J/\Psi}-M_{\eta_c}$ compared to the experimental value. The errors in the plot are purely statistical and errors from the tuning of the charm quark mass are non-negligible\footnote{Please refer to Section \ref{hl_results} for an estimate of this uncertainty in the case of the $D_s$ hyperfine-splitting.}. Within our approach, the hyperfine-splitting is expected to show significant discretization errors \cite{Burch:2009az} which we can not further investigate, as the library of PACS-CS lattices used contains only one lattice spacing. This result indicates that discretization effects of a similar magnitude may be expected in other observables. In addition, disconnected diagrams which are neglected in our simulation will have a small effect which may be non-negligible for the hyperfine-splittings. In \cite{Levkova:2010ft} these contributions to the charmonium hyperfine-splitting have been calculated for a reasonable fitting Ansatz. They are found to be negative and therefore can not explain the difference of our results compared to experimental values. Our results also show a small but statistically significant dependence on the strange sea quark mass. Notice however that our charm quark tuning was performed for the ensemble with lightest sea quark mass and $\kappa_s=0.1364$. For all other charmonium mass differences the dependence on the strange sea quark mass is very mild and there is no statistically significant difference between the two ensembles at $\kappa_{u/d}=0.13754$.

\begin{figure}[!tbh]
\includegraphics[width=85mm,clip]{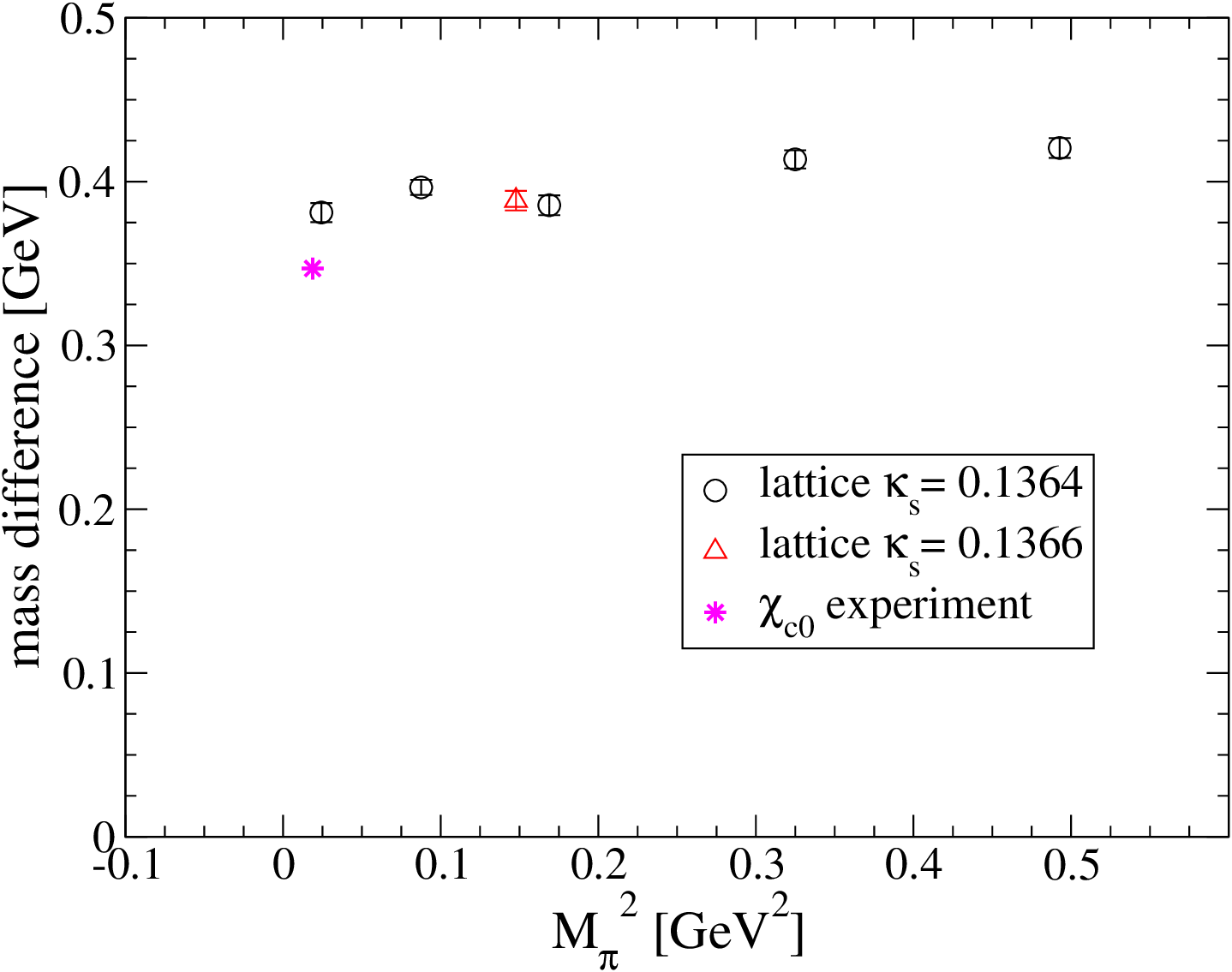}\\
\includegraphics[width=85mm,clip]{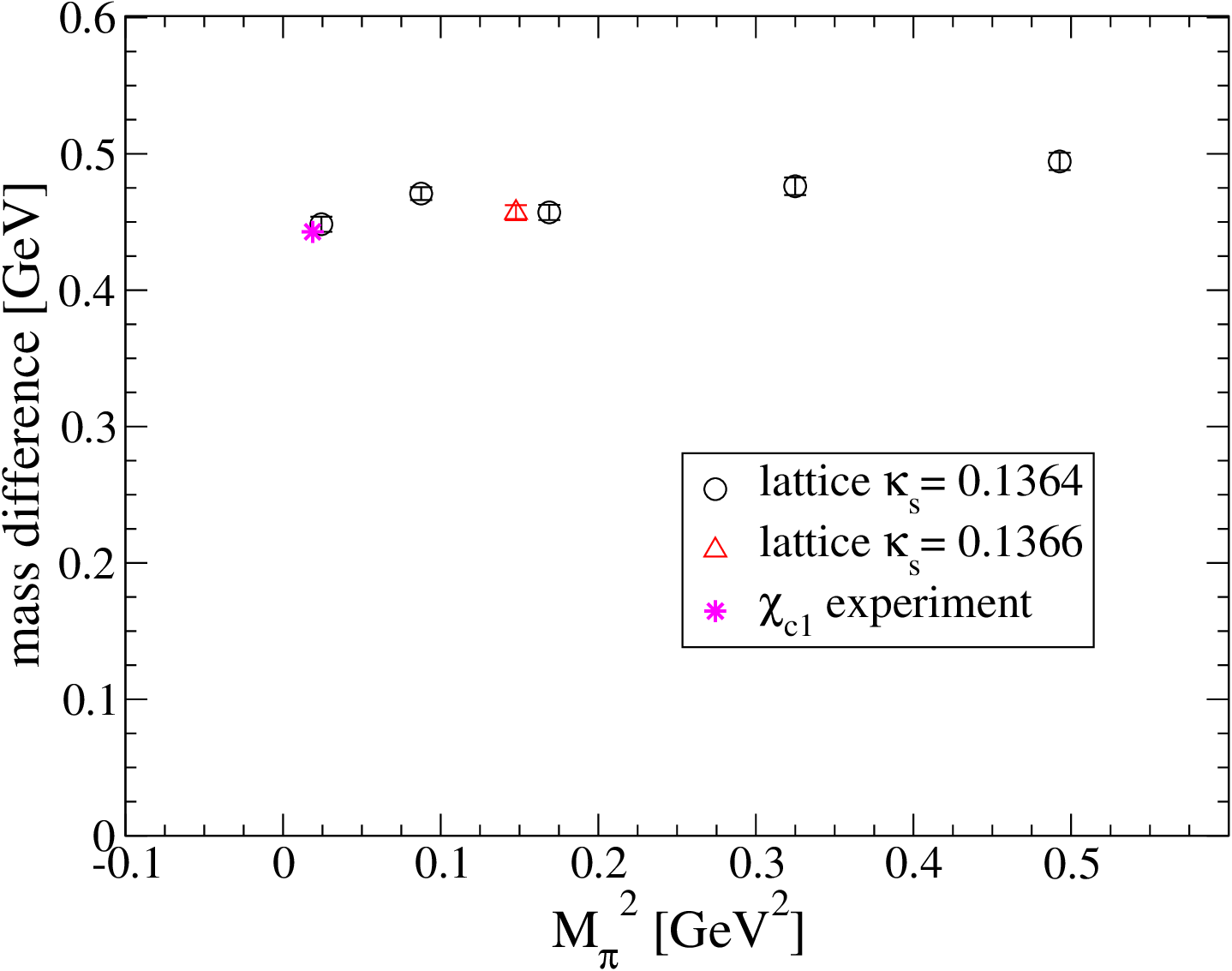}\\
\includegraphics[width=85mm,clip]{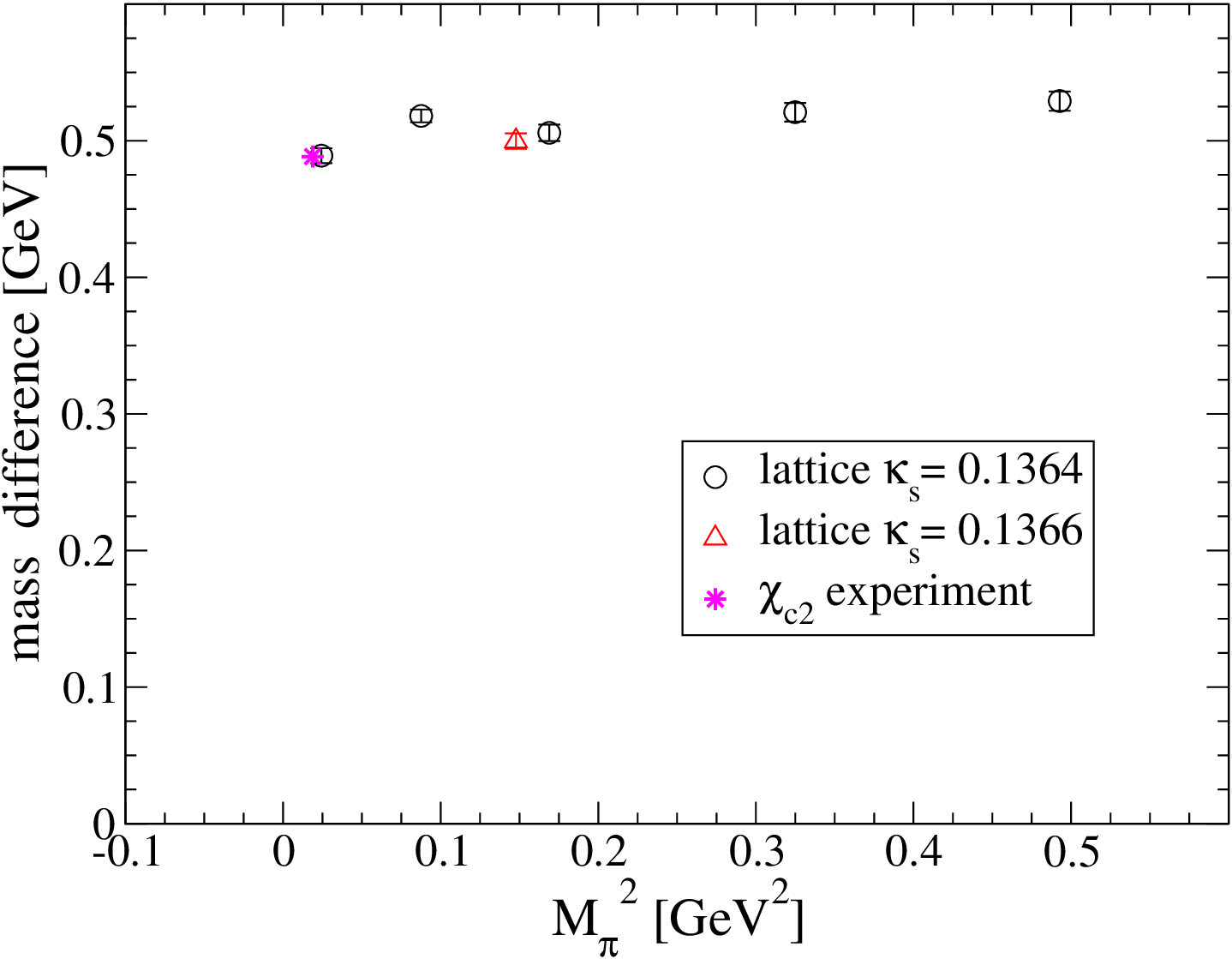}
\caption{Chiral behavior for the single charmonium P-wave states compared to experiment. As for all plots, the numbers are mass differences with respect to the spin-averaged ground state. The error displayed is statistical only. The uncertainty in the lattice scale implies that all points could be moved uniformly by $\approx$1.4\%.} 
\label{hh_pwaves}
\end{figure}

\begin{figure}[t]
\includegraphics[width=85mm,clip]{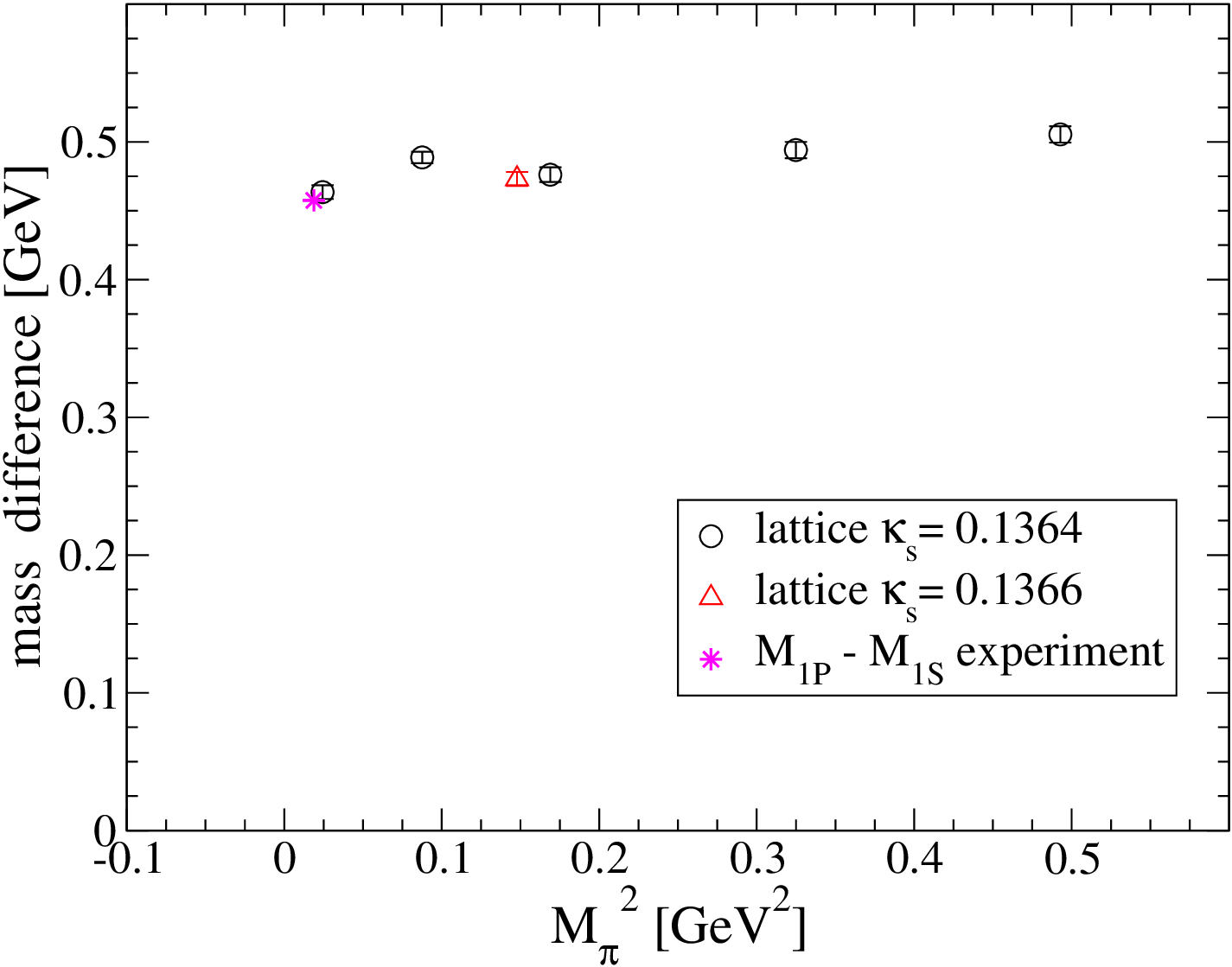}
\caption{Splitting $M_{\overline{1P}}-M_{\overline{1S}}$ between the spin-averaged $1P$ triplet state and the spin-averaged ground state. The error displayed is statistical only. The uncertainty in the lattice scale implies that all points could be moved uniformly by $\approx$1.4\%.} 
\label{hh_1S_1P}
\end{figure} 

Next consider the spin-triplet P-wave states, which are the ground states with quantum numbers $0^{++}$, $1^{++}$ and $2^{++}$ corresponding to the $\chi_{c0}$, $\chi_{c1}$ and $\chi_{c2}$. Figure \ref{hh_pwaves} shows the difference between the masses of the individual states and the spin-averaged ground state mass for all ensembles. At the lowest quark mass the ground state P-waves calculated at finite lattice spacing show good qualitative agreement with experiment, with the $\chi_{c0}$ ground state showing the largest discrepancy. While our results at the three heaviest and at the lowest quark masses suggest that the chiral extrapolation is within errors well described by the leading order term linear in the pion mass squared, the data from the ensemble at $\kappa_{u/d}=0.13770$ deviates from this behavior. This peculiarity is found in most of the observables we investigate. Note that a similar behavior can be seen in \cite{Mahbub:2010rm} where the first positive parity excitation of the Nucleon, the \emph{Roper resonance} is calculated on the same gauge configurations. In \cite{Mahbub:2010rm} this is interpreted as ``significant chiral curvature''. Our data are not suggestive of this explanation. Without additional simulation data we refrain from interpreting this behavior and will instead quote as our final results those at the lightest pion mass $m_\pi\approx 156\mathrm{MeV}$. The effect of the remaining small chiral extrapolation would be negligible compared to current statistical and other systematical uncertainties.

\begin{figure}[t]
\includegraphics[width=85mm,clip]{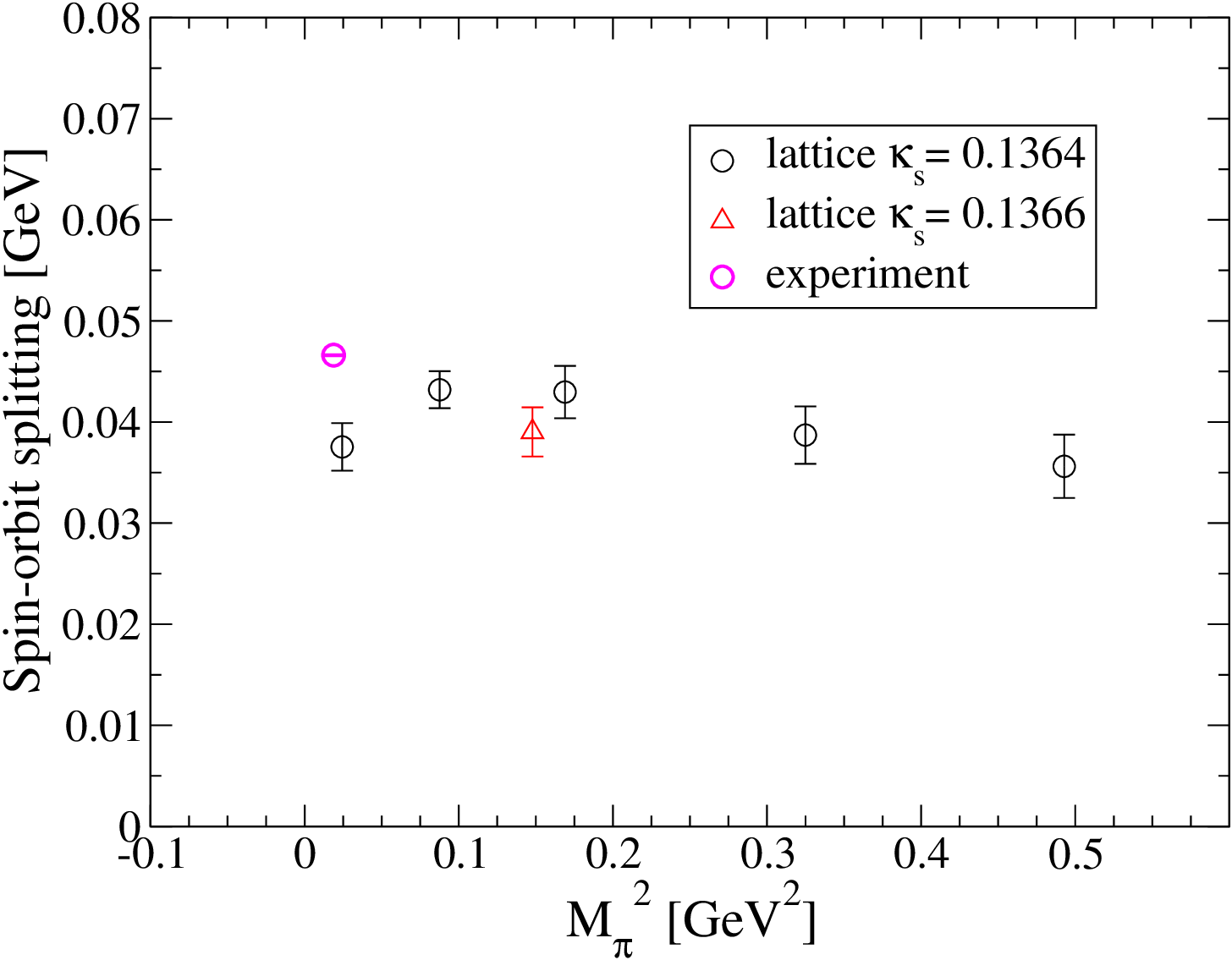}
\includegraphics[width=85mm,clip]{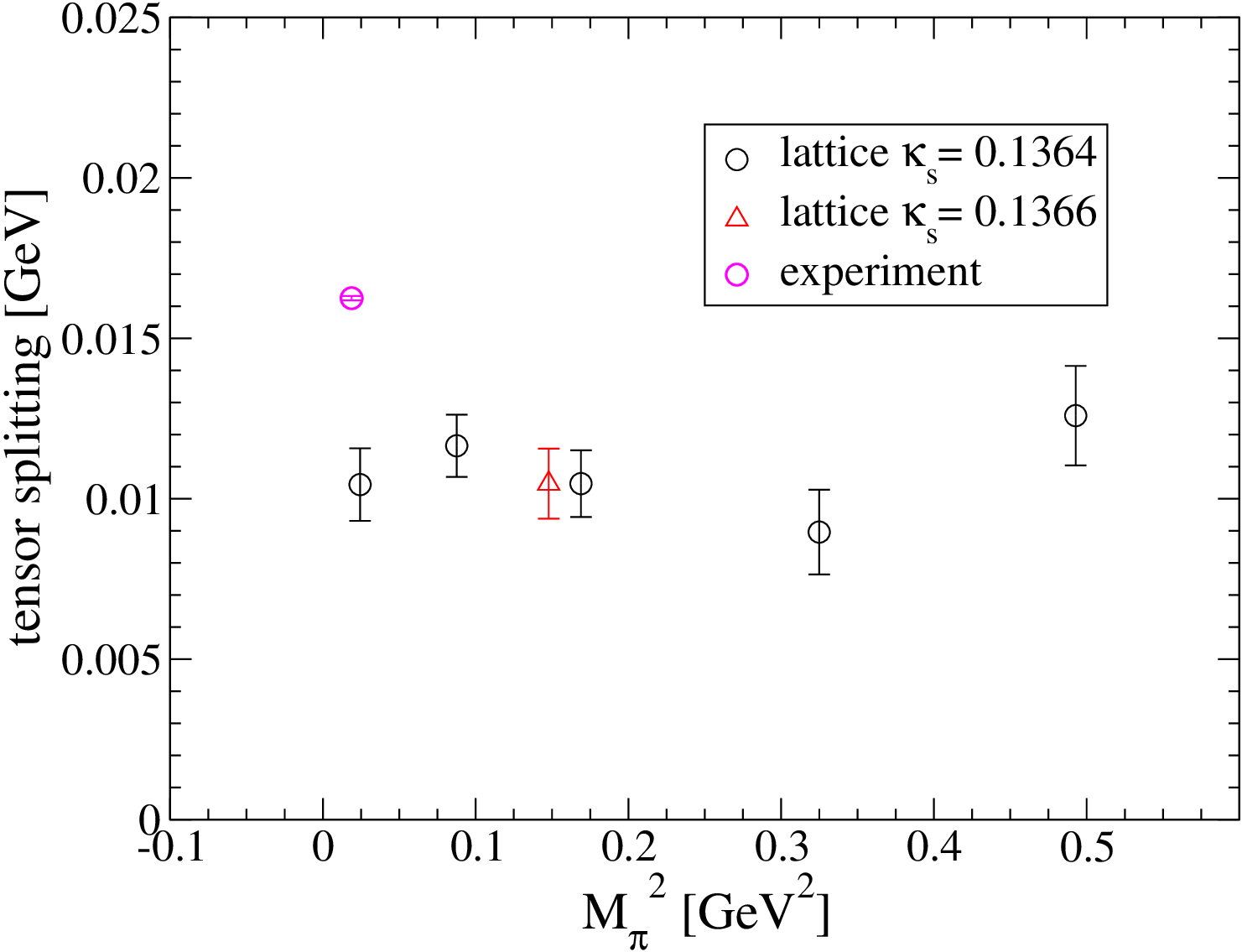}
\caption{ Spin-orbit splitting (upper panel) and tensor splitting (lower panel) for the triplet of 1P states. The error does not include the error due to the tuning of the charm-quark hopping parameter $\kappa_c$ and is statistical only. The uncertainty in the lattice scale implies that all points could be moved uniformly by $\approx$1.4\%.} 
\label{spin-dependant}
\end{figure}

In addition, we also form the combinations (see \cite{Bernard:2010fr} and references therein)
\begin{align}
M_{\overline{1P}}&=\frac{1}{9}(M_{\chi_{c0}}+3M_{\chi_{c1}}+5M_{\chi_{c2}}) ,\\
M_{Spin-Orbit}&=\frac{1}{9}(5M_{\chi_{c2}}-3M_{\chi_{c1}}-2M_{\chi_{c0}}) ,\\
M_{Tensor}&=\frac{1}{9}(3M_{\chi_{c1}}-M_{\chi_{c2}}-2M_{\chi_{c0}}) .
\end{align}
$M_{\overline{1P}}$ is the Spin-averaged mass of the P wave triplet. $M_{Spin-Orbit}$ is sensitive to the spin-orbit interaction while $M_{Tensor}$ is sensitive to the spin-spin interaction \cite{Bernard:2010fr}. Figure \ref{hh_1S_1P} shows the splitting between the spin-averaged ground state mass $M_{\overline{1S}}$ and $M_{\overline{1P}}$. While we are unable to perform a continuum extrapolation and small discrepancies to experiment are expected, the figure suggests that the spin-averaged quantities are well reproduced within our approach.

The results for $M_{Spin-Orbit}$ and $M_{Tensor}$ are plotted in Figure \ref{spin-dependant}. The upper panel shows the Spin-Orbit splitting of the 1P states, which is slightly underestimated in our simulation, similar to the hyperfine-splitting. The lower panel shows the tensor splitting of the 1P states. Compared to experiment we obtain the correct sign and a similar magnitude with the value off about 30\%.

\begin{figure}[t]
\includegraphics[width=85mm,clip]{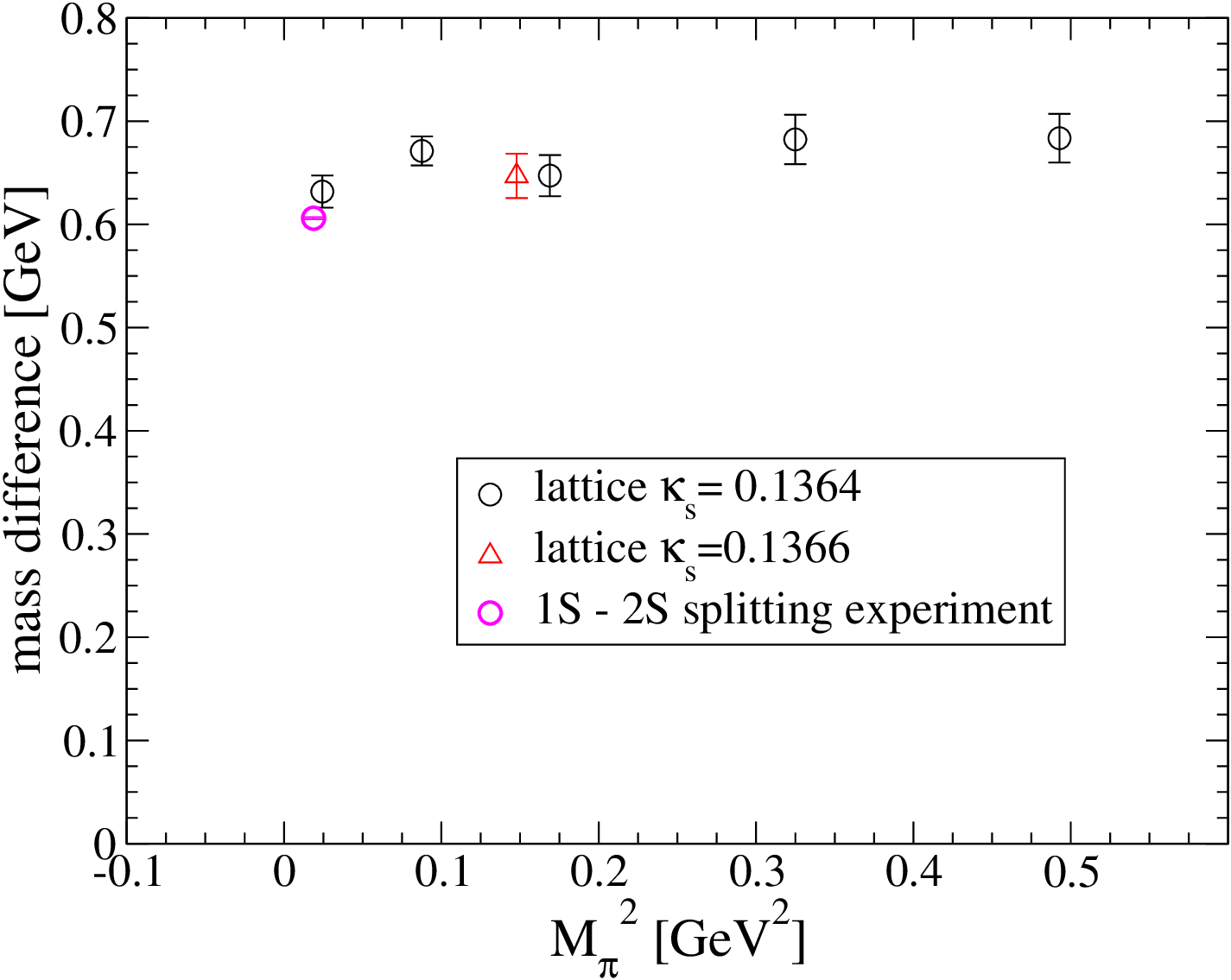}
\caption{Splitting $M_{\overline{2S}}-M_{\overline{1S}}$ between the 2S and 1S spin-averaged states. The error displayed is statistical only. The uncertainty in the lattice scale implies that all points could be moved uniformly by $\approx$1.4\%.} 
\label{hh_1S_2S}
\end{figure}

So far the discussion was restricted to properties of ground states. In the $1^{--}$ and $0^{-+}$ channels our basis allows for a determination of excited energy levels. While these states are more noisy than the ground states and the fit ranges (see tables in the Appendix \ref{tables_b}) are more limited, at least the 2S states are well determined from our charmonium data. Figure \ref{hh_1S_2S} shows the splitting between the 2S and 1S spin-averaged states. Again the result agrees qualitatively with experiment suggesting that our approach is well suited to reproduce spin-averaged quantities.

\begin{figure}[tb]
\includegraphics[width=85mm,clip]{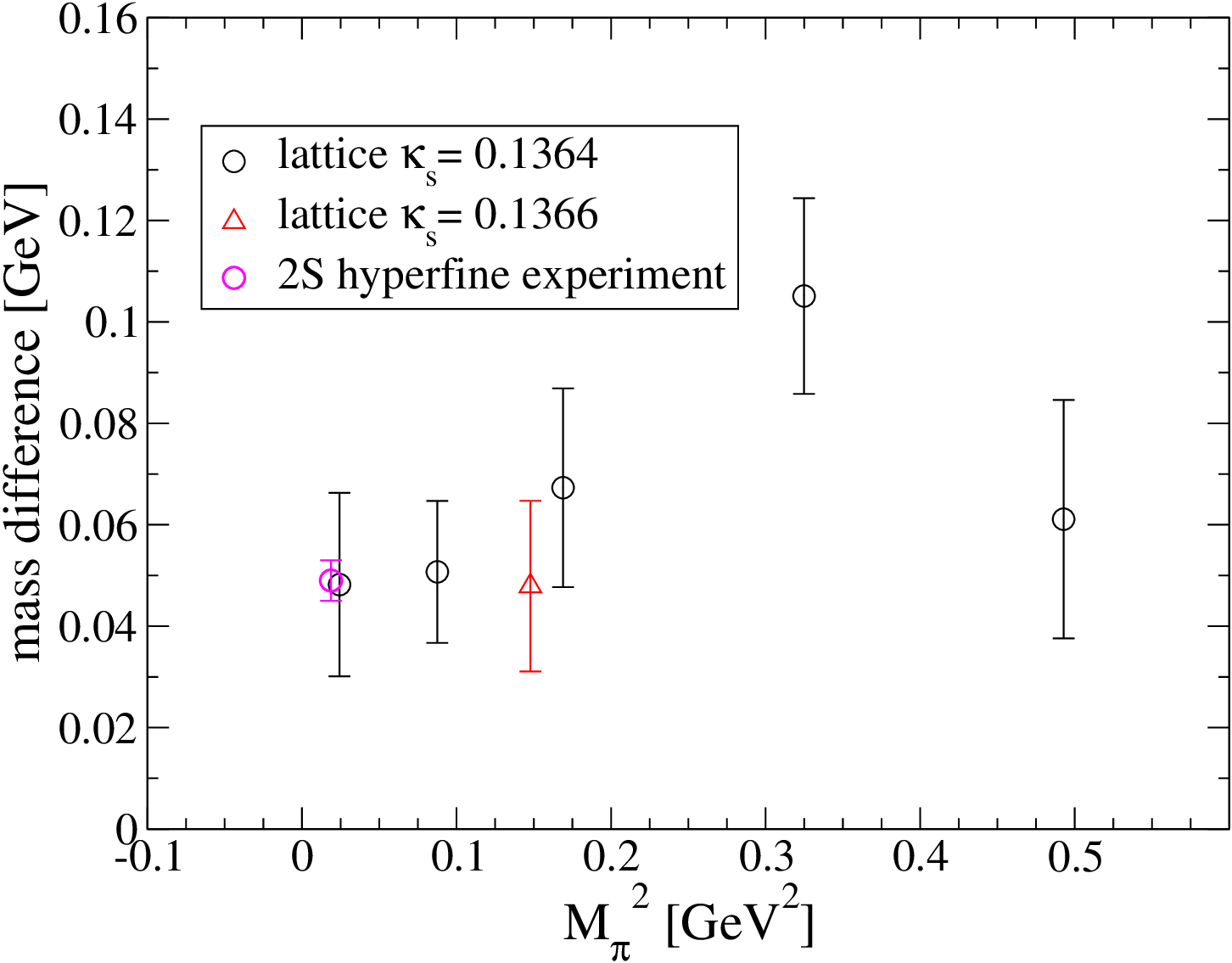}
\caption{Hyperfine-splitting for the 2S charmonium states compared to the physical splitting. The error is statistical only.} 
\label{hh_2S_hyperfine}
\end{figure}

Figure \ref{hh_2S_hyperfine} shows the 2S hyperfine-splitting. Unlike the 1S hyperfine-splitting this quantity is not determined very accurately from our data. Just like in experiment the 2S hyperfine-splitting is smaller than the 1S hyperfine-splitting. This again gives us confidence that we capture the important physics well. This concludes the discussion of charmonium results. Final numbers from the ensemble with lightest sea quarks can be found in Section \ref{conclusion} along with a table that summarizes the results and their respective uncertainties. A table with charmonium energy levels can be found in Appendix \ref{tables_b}.
 
\section{\label{hl_results}Heavy-light mesons}

In the heavy quark limit, the spectrum of S- and P-wave heavy-light mesons can be classified by the total angular momentum $j$ of the light degrees of freedom\footnote{Note that we use small $j$ when referring to the heavy quark limit and $J$ when referring to the Spin away from the heavy quark limit.}. This classification contains one S-wave doublet with $j^P=\frac{1}{2}^-$ containing the $D$ and $D^\star$ mesons in case of the heavy-light states and the $D_s$ and $D_s^\star$ mesons in case of the heavy-strange states. The P-wave states fall into two multiplets, one with $j^P=\frac{1}{2}^+$ and one with $j^P=\frac{3}{2}^+$. Taking a look at the experimental situation one would identify the charmed-strange mesons $D_{s0}^\star(2317)$ and $D_{s1}(2460)$ as the charmed-strange version of the P-wave heavy quark doublet with $j=\frac{1}{2}$ and the  $D_{s1}(2536)$ and the  $D_{s2}^\star(2573)$ as the charmed-strange version of the multiplet with $j=\frac{3}{2}$. While the mass of states corresponding to $j^P=\frac{1}{2}^-$ and $j^P=\frac{3}{2}^+$ states in the heavy mass limit is quite close to the values expected from the quark model \cite{Godfrey:1985xj}, the  states corresponding to the heavy-strange $j^P=\frac{1}{2}^+$ doublet are substantially lighter \cite{0954-3899-37-7A-075021}, putting them below the $D+K$ and $D^\star+K$ thresholds. Even more puzzling however is a comparison to the spectrum of heavy-light $D$ mesons, where similar multiplets are expected in the heavy quark limit. In particular, the charmed-light version of the $j^P=\frac{1}{2}^+$ doublet is given by the $D_0^\star(2400)$ and the $D_1(2420)$. This corresponds to a mass-splitting between these states and the spin-averaged S-wave states which is much larger for $D$ mesons than for $D_s$ mesons. Such a behavior is not expected in simple potential models. Furthermore, chiral perturbation theory for static-light mesons suggests the opposite \cite{Becirevic:2004uv}.

Before proceeding to report results, let us discuss previous lattice calculations of charmed and charmed-strange mesons focusing in particular on results for the $j^P=\frac{1}{2}^+$ doublet. Lewis and Woloshyn \cite{Lewis:2000sv} calculated the spectrum of  S- and P-wave heavy-light mesons in lattice NRQCD using quenched gauge configurations. At the time of this study, the $J^P=0^+$ states were not yet measured. While their results for the heavier $j^P=\frac{3}{2}$ doublet agree within errors with the experimental results, the $D_0^\star$, $D_{s0}^\star$ and the  $j^P=\frac{1}{2}$ $D_{s1}$ were predicted at masses much larger than the experimental masses. Another early calculation using lattice NRQCD was performed by Hein et al. \cite{Hein:2000qu}. In \cite{Boyle:1997aq,Boyle:1997rk} the UKQCD collaboration reports on results using a relativistic clover-type action based on the Fermilab method \cite{ElKhadra:1996mp}. Results from the studies \cite{Hein:2000qu,Boyle:1997aq,Boyle:1997rk} along with a calculation in the static limit on $n_f=2$ dynamic gauge configurations have been presented by Bali \cite{Bali:2003jv}. All these results share the common feature that the ground state in the $J^P=0^+$ channel is estimated at a mass exceeding the mass of the experimental $D_{s0}^\star(2317)$ by roughly 130-200 MeV. Furthermore, where calculated, the splitting between the S-wave states and the $0^+$ P-wave states for heavy-strange mesons was determined to be larger or of the same size than for heavy-light states.

\begin{figure}[t]
\includegraphics[width=85mm,clip]{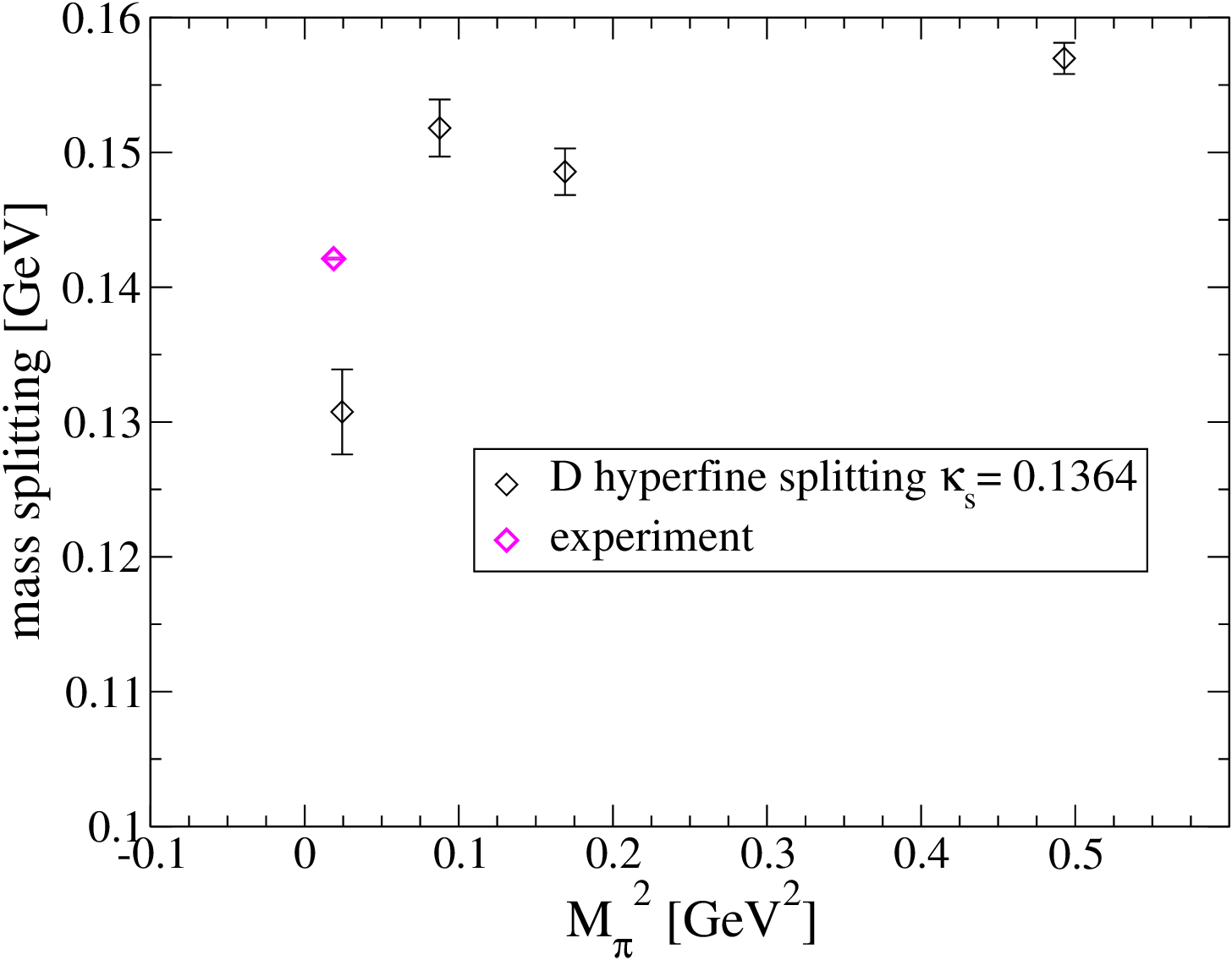}\\
\includegraphics[width=85mm,clip]{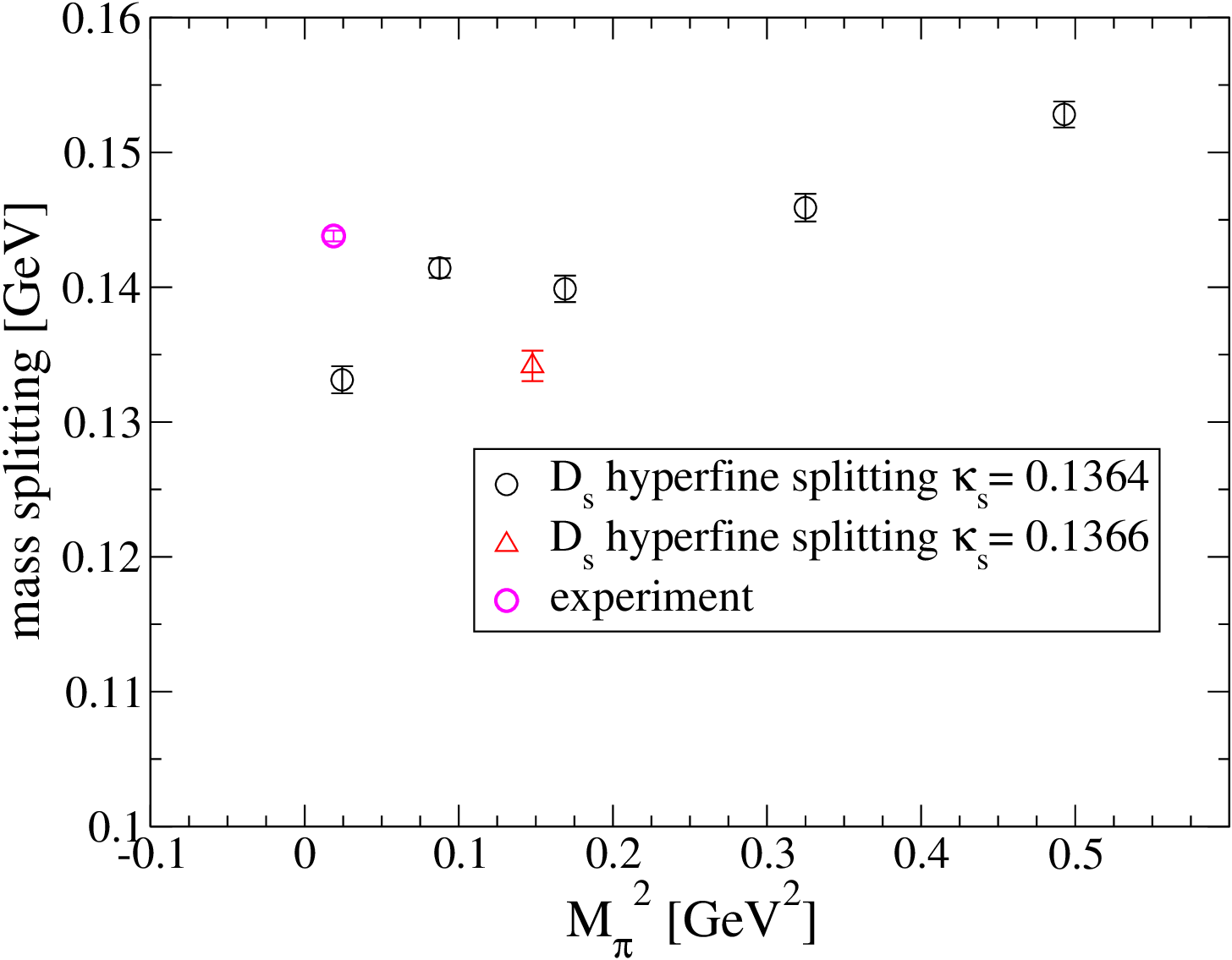}\\
\caption{Hyperfine-splitting for heavy-light (upper panel) and heavy-strange (lower panel) mesons compared to the physical splitting. The errors are statistical only and the leading discrepancy is caused by the lack of a continuum extrapolation. For further discussion, please refer to the text.} 
\label{hl_hyperfine}
\end{figure}

In addition to the above simulations, the UKQCD collaboration also presented results from both quenched and one dynamical ensemble in \cite{Dougall:2003hv}. While they report a somewhat smaller discrepancy with respect to experiment, their scale setting results in a quite large value of $0.55\mathrm{fm}$ for the Sommer scale $r_0$. Typical values for the Sommer scale from recent dynamical simulations however suggest a significantly smaller value $r_0\approx0.46\dots0.49\mathrm{fm}$. In particular the scale setting for our simulations, as determined by the PACS-CS collaboration, corresponds to $r_0=0.4921(64)(+74)(-2)$ \cite{Aoki:2008sm}. A smaller value of $r_0$ results in larger mass splittings. The scale setting itself is intimately related to the determination of the physical point, which again emphasizes the need for light dynamical quarks.

More recently, the $\chi$QCD collaboration presented preliminary results from both quenched \cite{Dong:2008vd} and dynamical \cite{Dong:2009wk} lattices using relativistic quarks and overlap valence fermions. Their initial results for both quenched and dynamical simulations suggest good agreement between lattice results and the low-lying charmed-strange mesons.

The rest of this section presents results for the spectrum of heavy-strange and heavy-light mesons. The final numbers and an overview of the results can be found in Section \ref{conclusion}.

\begin{figure*}[p]
\includegraphics[height=70mm,clip]{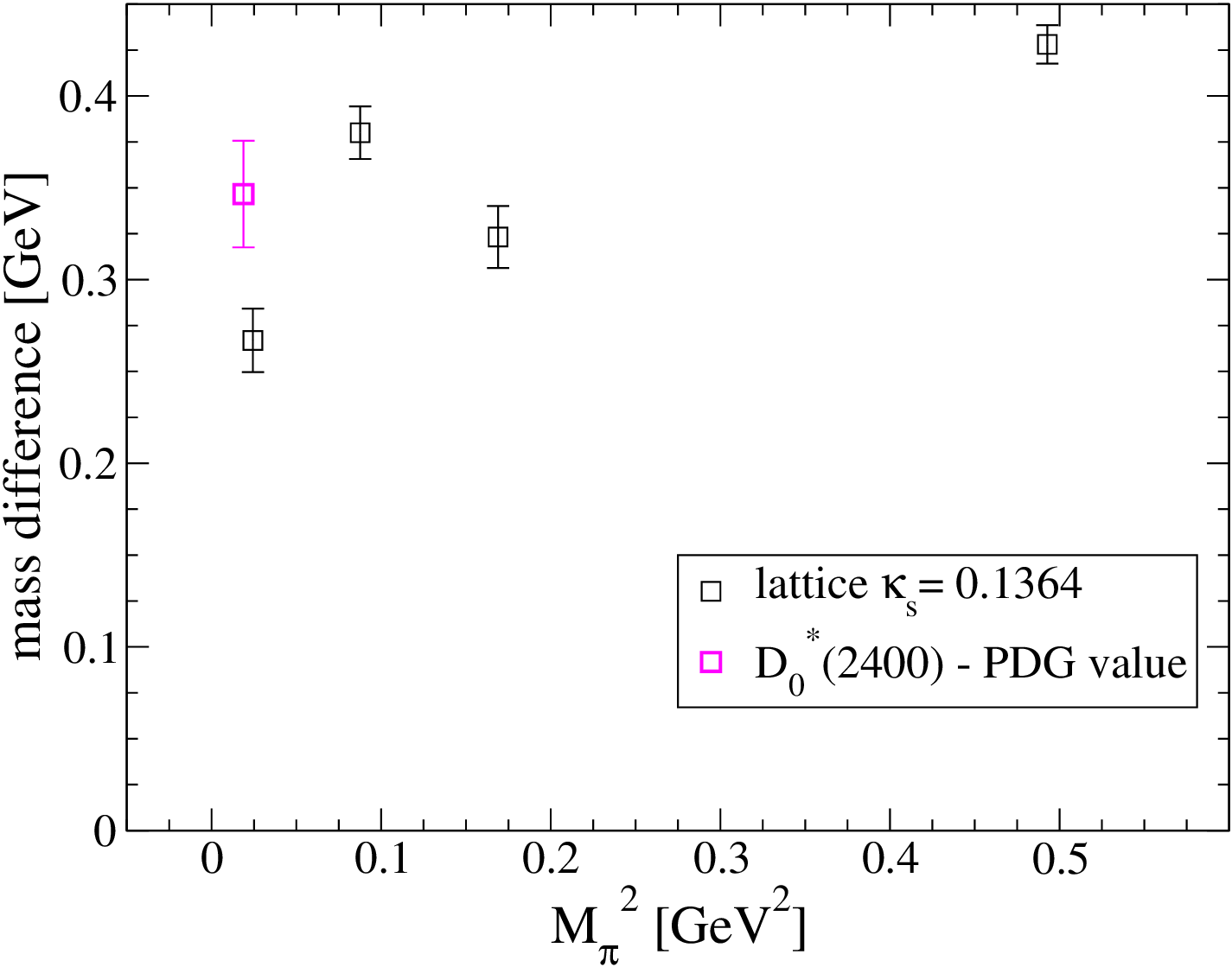}
\includegraphics[height=70mm,clip]{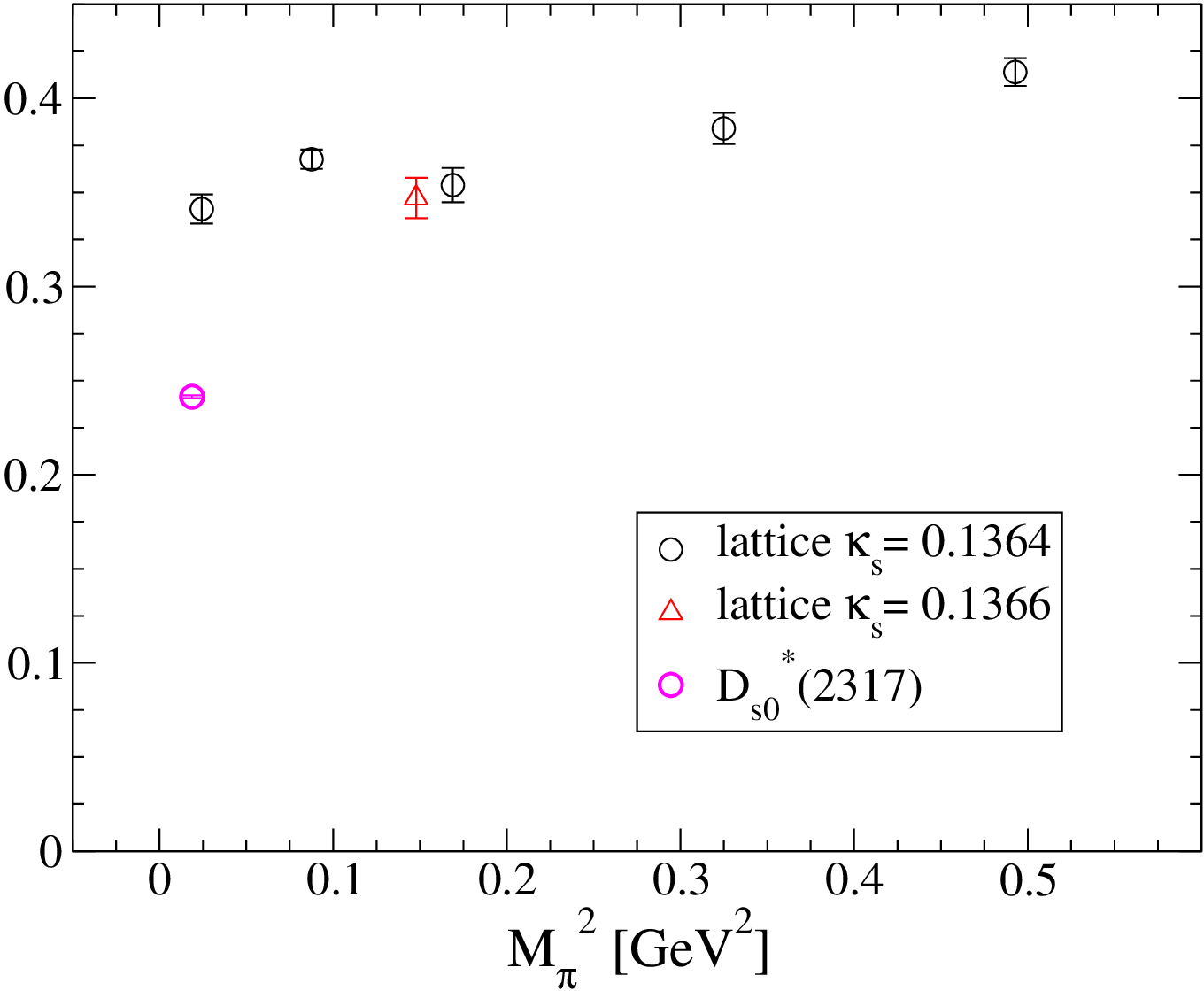}\\
\includegraphics[height=70mm,clip]{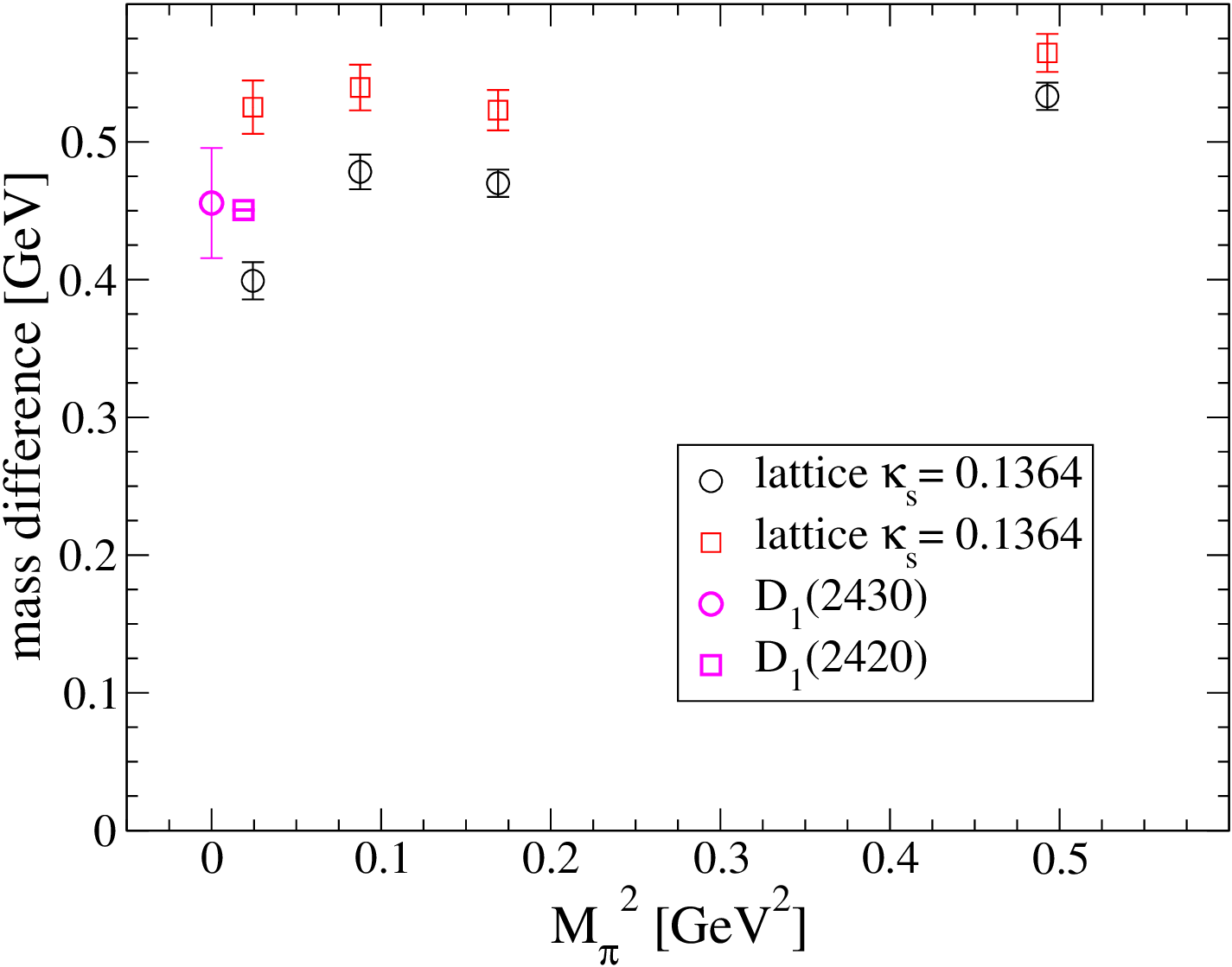}
\includegraphics[height=70mm,clip]{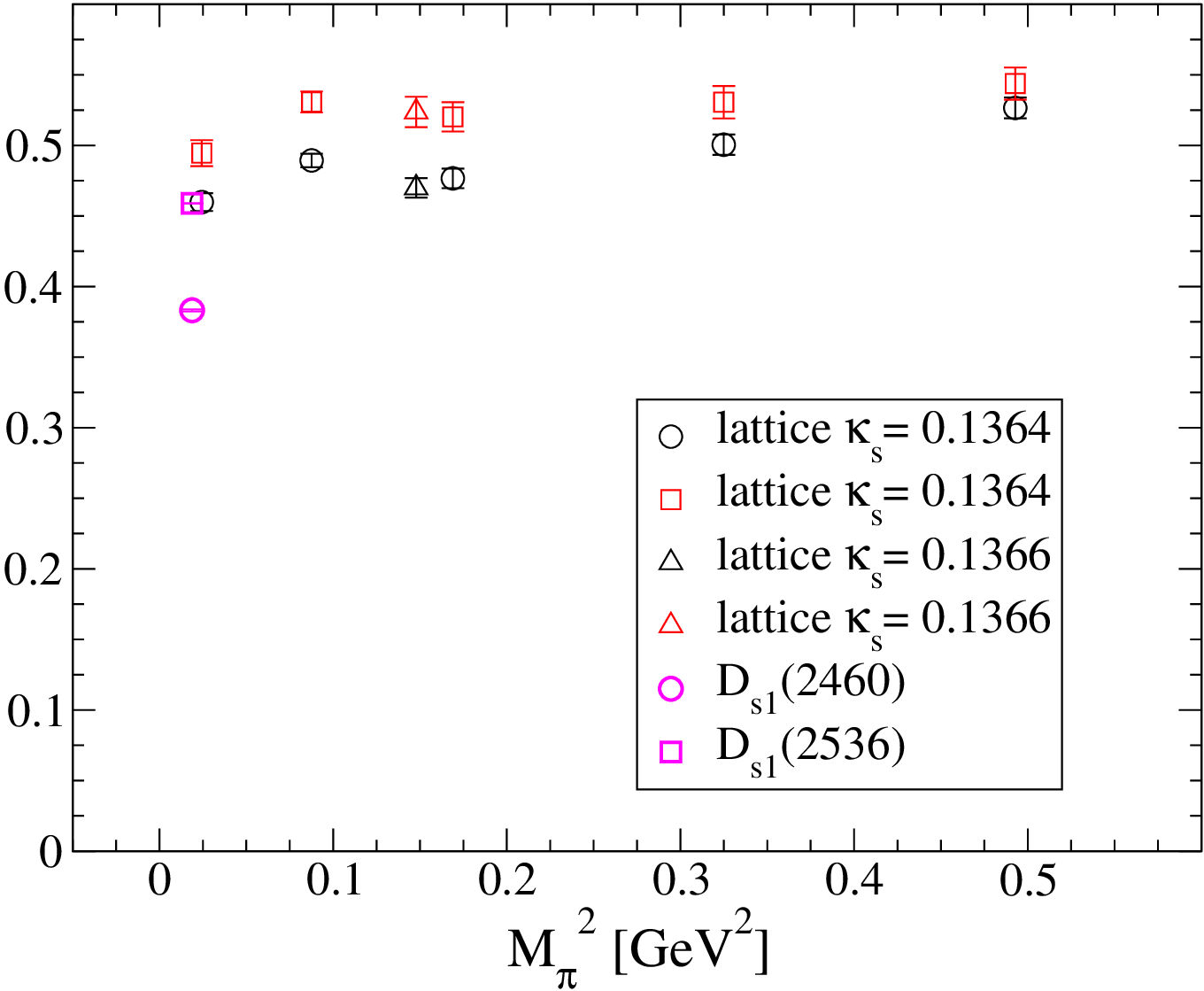}\\
\includegraphics[height=70mm,clip]{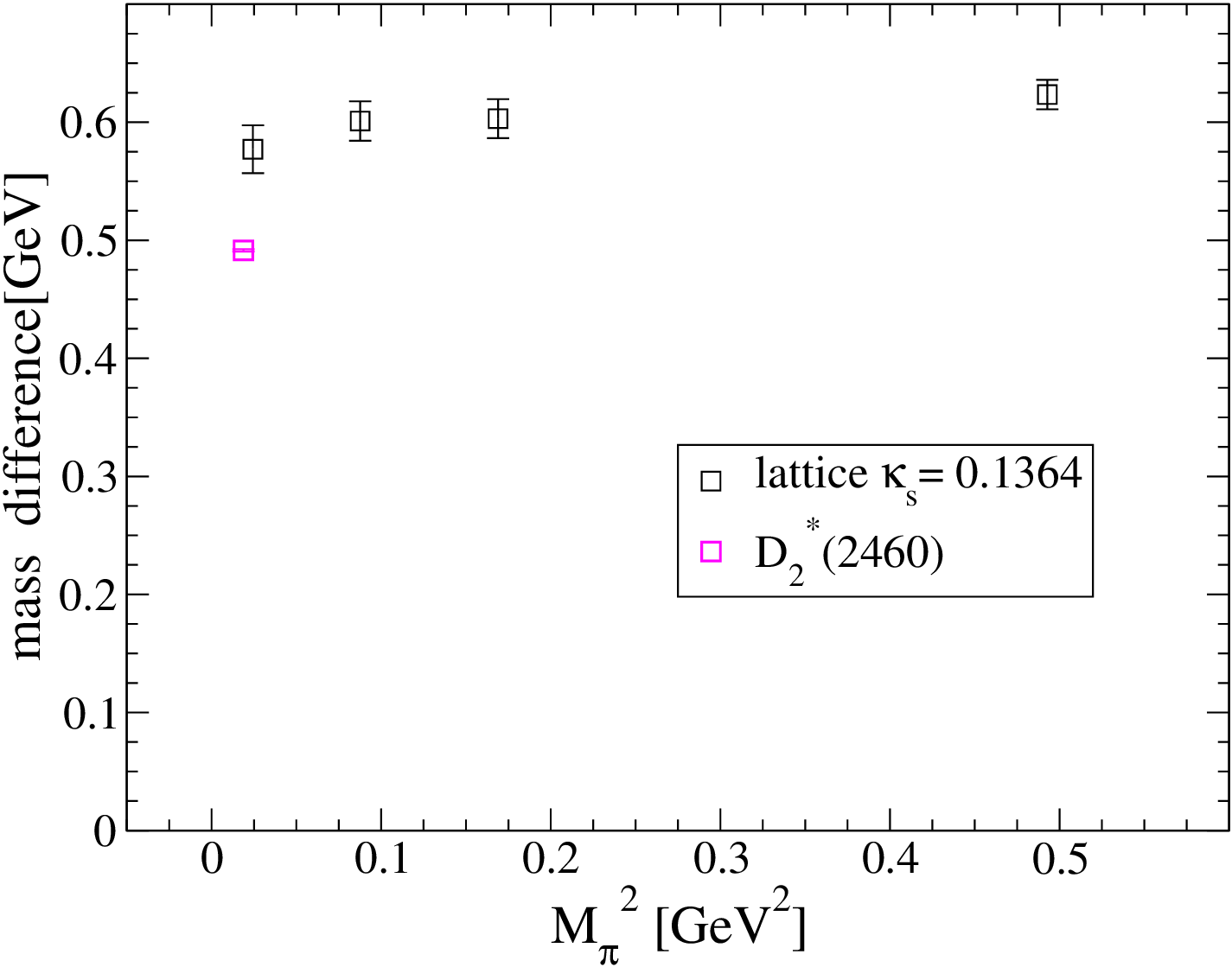}
\includegraphics[height=70mm,clip]{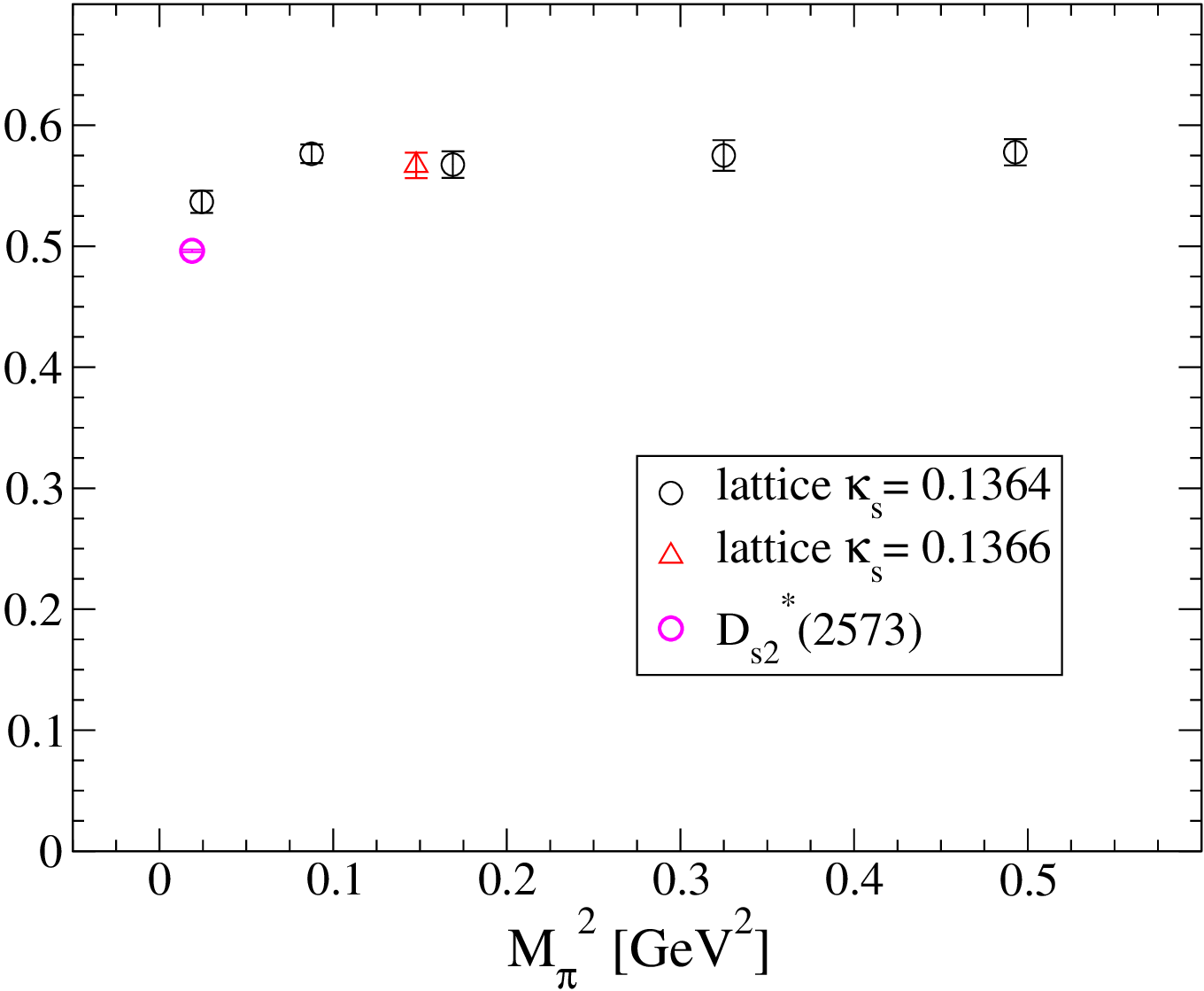}
\caption{Chiral behavior for the $D$ (left panels) and $D_s$ (right panels) P-wave states compared to experiment. The $D(2430)$ (left middle panel) has been slightly offset from the physical pion mass to improve visibility. As for all plots, the numbers are mass differences with respect to the respective spin-averaged ground state. The error displayed is statistical only. The uncertainty in the lattice scale implies that all points could be moved uniformly by $\approx$1.4\%.} 
\label{hl_pwaves}
\end{figure*}

Figure \ref{hl_hyperfine} shows the hyperfine-splittings for the $D$ (upper panel) and $D_s$ (lower panel) mesons. Similar to Charmonium, the hyperfine-splitting turns out to be somewhat to small, which we attribute partly to non-negligible discretization effects. This discrepancy is somewhat smaller for heavy-light states than for charmonium. Like for charmonium the hyperfine-splitting on the ensemble with $\kappa_s=0.1366$ is slightly smaller than for the other ensembles. Another non-negligible source of error is the uncertainty in the determination of $\kappa_c$. From the tuning runs, this uncertainty is estimated to be roughly 3.2 MeV for $D_s$ mesons and we assume a similar dependence for $D$ mesons. In the case of $D_s$ mesons where statistical uncertainties are small, neglecting the data point at our second lowest quark mass, the remaining data could be fit with a good $\chi^2/\mathrm{d.o.f.}$ by a fit linear in the pion mass squared. Although the results have much larger statistical errors, chiral effects seem to be more significant for $D$ mesons. 

\begin{figure}[t]
\includegraphics[width=85mm,clip]{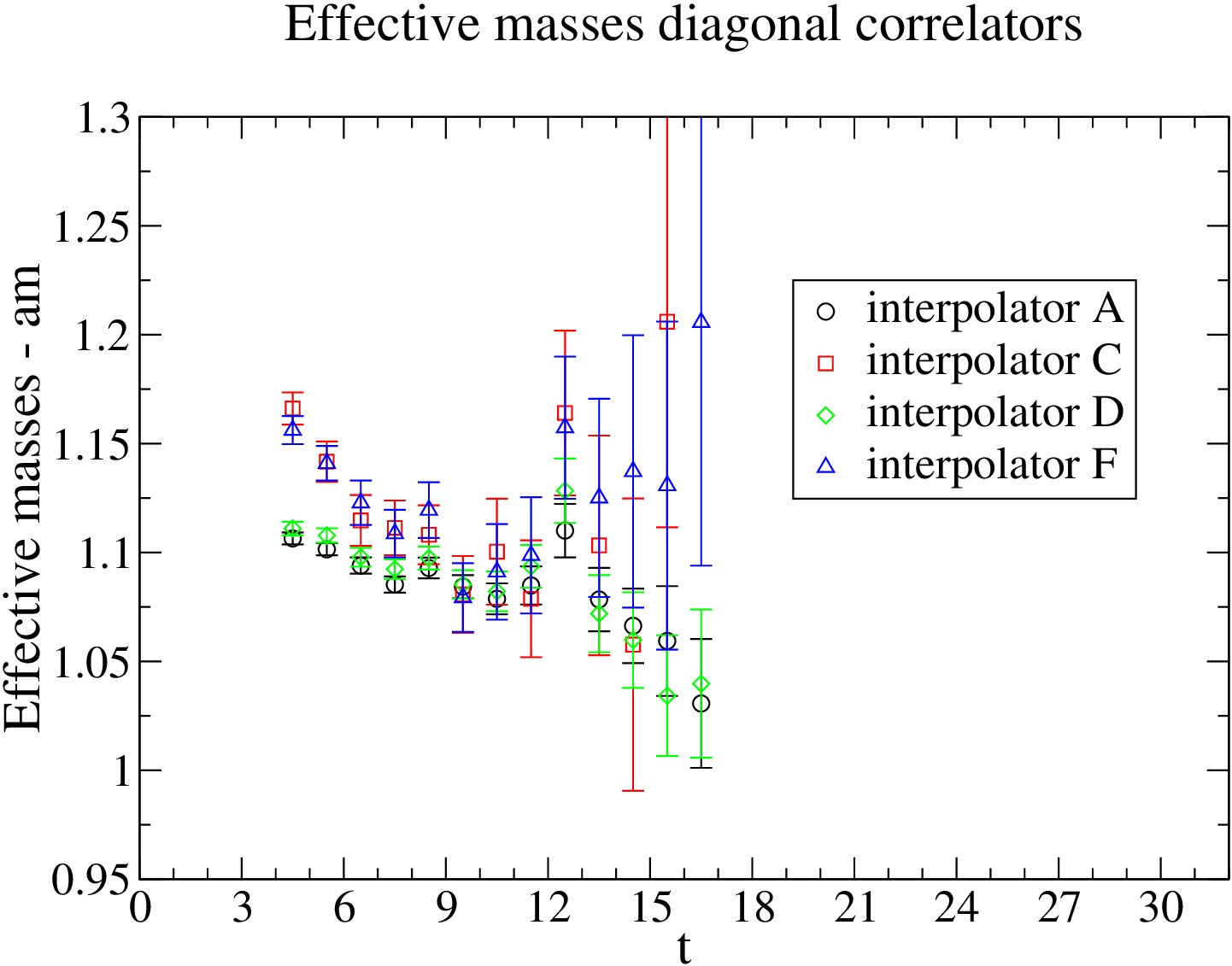}\\
\includegraphics[width=85mm,clip]{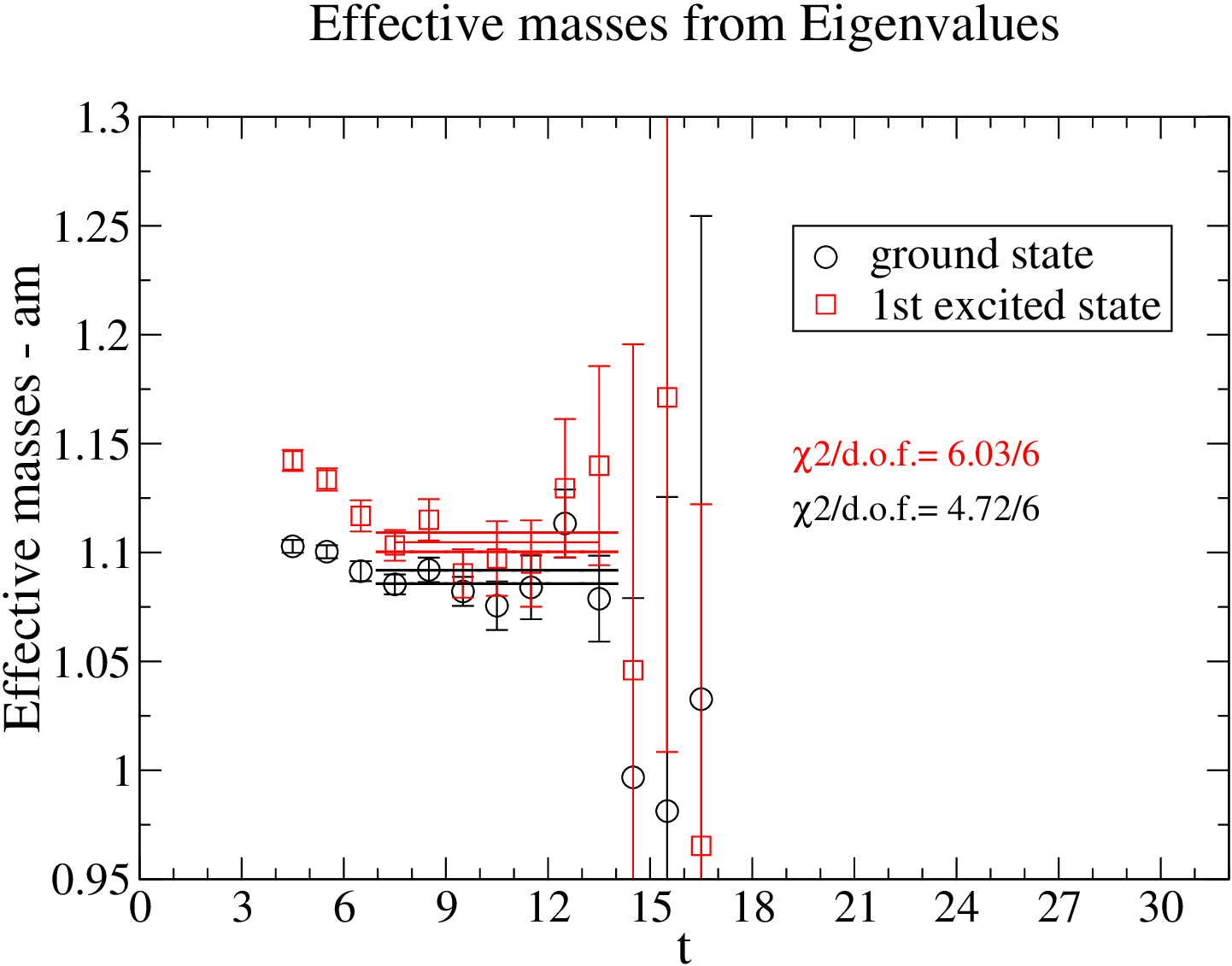}
\caption{Selected single diagonal correlators (upper panel) and lowest two states from a $4\times 4$ matrix of correlators (lower panel) in the $D_{s1}$ ($J^P=1^+$) channel. The data is from Ensemble 6. The interpolating fields are tabulated in Appendix \ref{source_tables}.} 
\label{1+_mixing}
\end{figure}

Figure \ref{hl_pwaves} shows results for the $D$ (left panels) and $D_s$ (right panels) P-wave states. As for charmonium, mass differences relative to the respective spin-averaged S-wave state are plotted. The top row shows the ground state in the $D_0^\star$ and $D_{s0}^\star$ ($J^P=0^+$) channels. In both cases, the mass splitting gets smaller as the light sea quark masses approach the chiral limit. Unlike suggested by experiment, the splitting for the $D_0^\star$ is smaller than for the $D_{s0}^\star$. While the discrepancy between our data and the $D_{s0}^\star(2317)$ is smaller for almost physical light quarks, calculation at light sea quarks alone can not resolve the apparent puzzle between the experimental results and the lattice data. For the $D$ mesons, statistical errors are larger. As previously remarked for the charmonium hyperfine-splitting in Section \ref{charmonium}, the data calculated on Ensemble 5 (see Table \ref{paratable}) seem to lead to larger mass splittings for all channels, while the rest of the data suggest that the leading order term proportional to the pion mass squared dominates the chiral extrapolation. This is especially evident for the $D_0^\star$ meson. For the final numbers and overview plots in Section \ref{conclusion}, we therefore quote the numbers obtained on the ensemble with the lightest pion mass. Any sensible functional form for a chiral extrapolation would alter these results by an amount much smaller than the dominant statistical and systematic errors.

The middle panels of Figure \ref{hl_pwaves} show the result for the two lowest states in the $D_1$ and $D_{s1}$ ($J^P=1^+$) channels. Both plots are from a matrix of correlators which includes interpolating fields corresponding to both positive and negative charge conjugation in the mass-degenerate case. In \cite{Bali:2003jv} the importance of this mixing has been emphasized, but the actual calculation neglected the mixing. In \cite{Boyle:1997rk} the mixing has been considered, but only the ground state could be resolved. Figure \ref{1+_mixing} illustrates the effect of this mixing. In particular the standard interpolators A and D (see Appendix \ref{tables_b}) both couple strongly to the ground state leading to a much smaller splitting than the mixed basis. The results from the $4\times4$ basis in Figure \ref{hl_pwaves} exhibit a mass splitting which is still somewhat small compared to the experimental splitting but enhanced compared to the analysis without mixing. For the $D_{s1}$ the results for the ground state lead to a splitting substantially larger than the $D_{s1}(2460)$, while the difference between the first excited state obtained in the calculation and the $D_{s1}(2536)$ is compatible with the non-negligible discretization errors one expects. For the $D_1$ mesons, the ground state displays a somewhat larger slope in the chiral extrapolation, while the splitting for the first excited state is comparable to the $D_{s1}$ case.

Finally the bottom two panels of Figure \ref{hl_pwaves} show the ground states of the $D_2^\star$ (left-hand side) and $D_{s2}^\star$ (right-hand side) with $J^P=2^+$. For some ensembles we encounter a systematic uncertainty (of a size similar to the purely statistical error) with regard to the choice of fit range in this channel. On the ensemble with lowest quark mass, the $D_{s2}^\star$ state is well determined though, and the difference from the experiment value is small. The $D_2^\star$ ground state is somewhat less well determined and the difference from experiment is of a similar magnitude as for its $D_1$ partner state in the $j^P=\frac{3}{2}^+$ heavy quark multiplet.

\begin{figure}[!]
\includegraphics[width=85mm,clip]{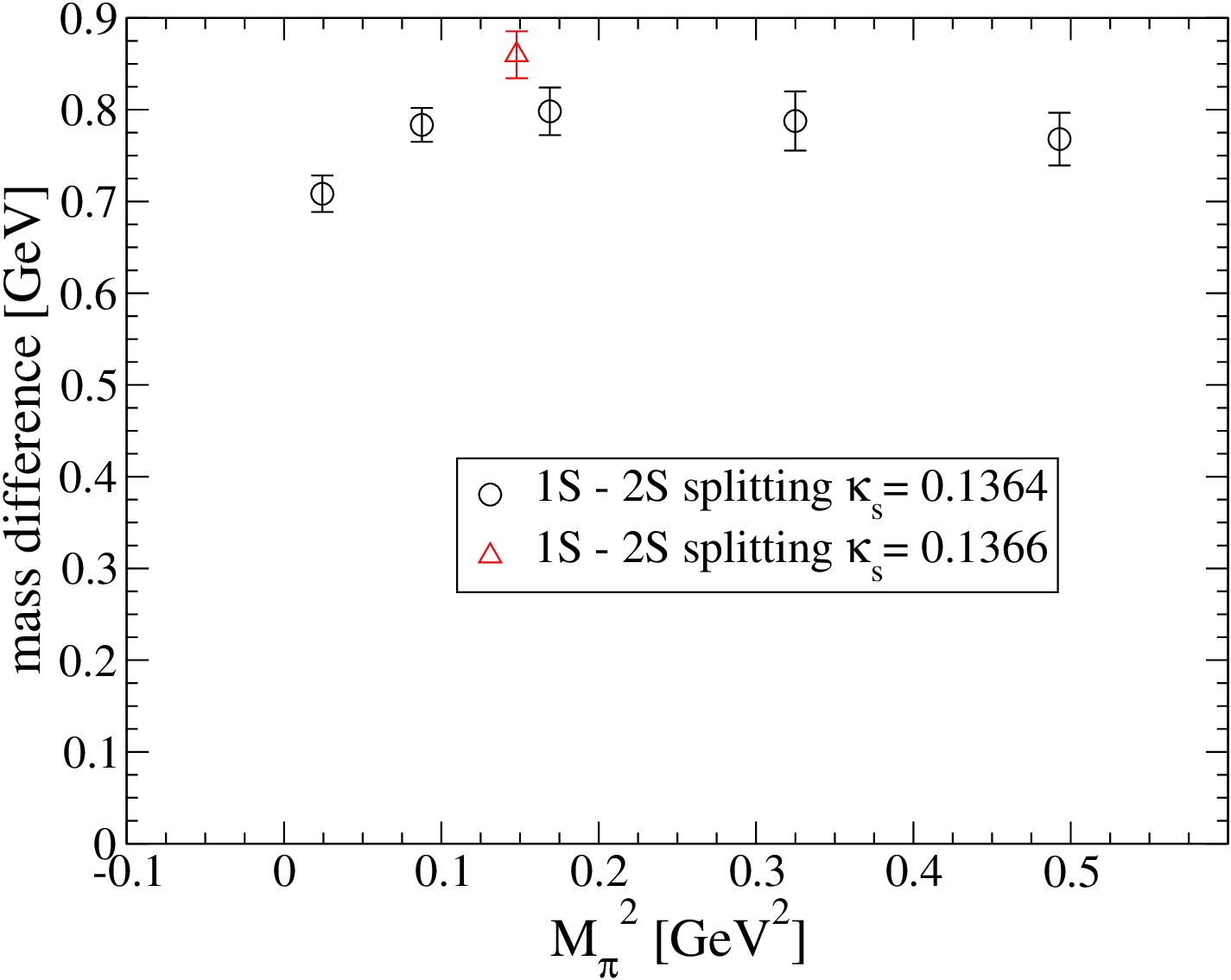}\\
\includegraphics[width=85mm,clip]{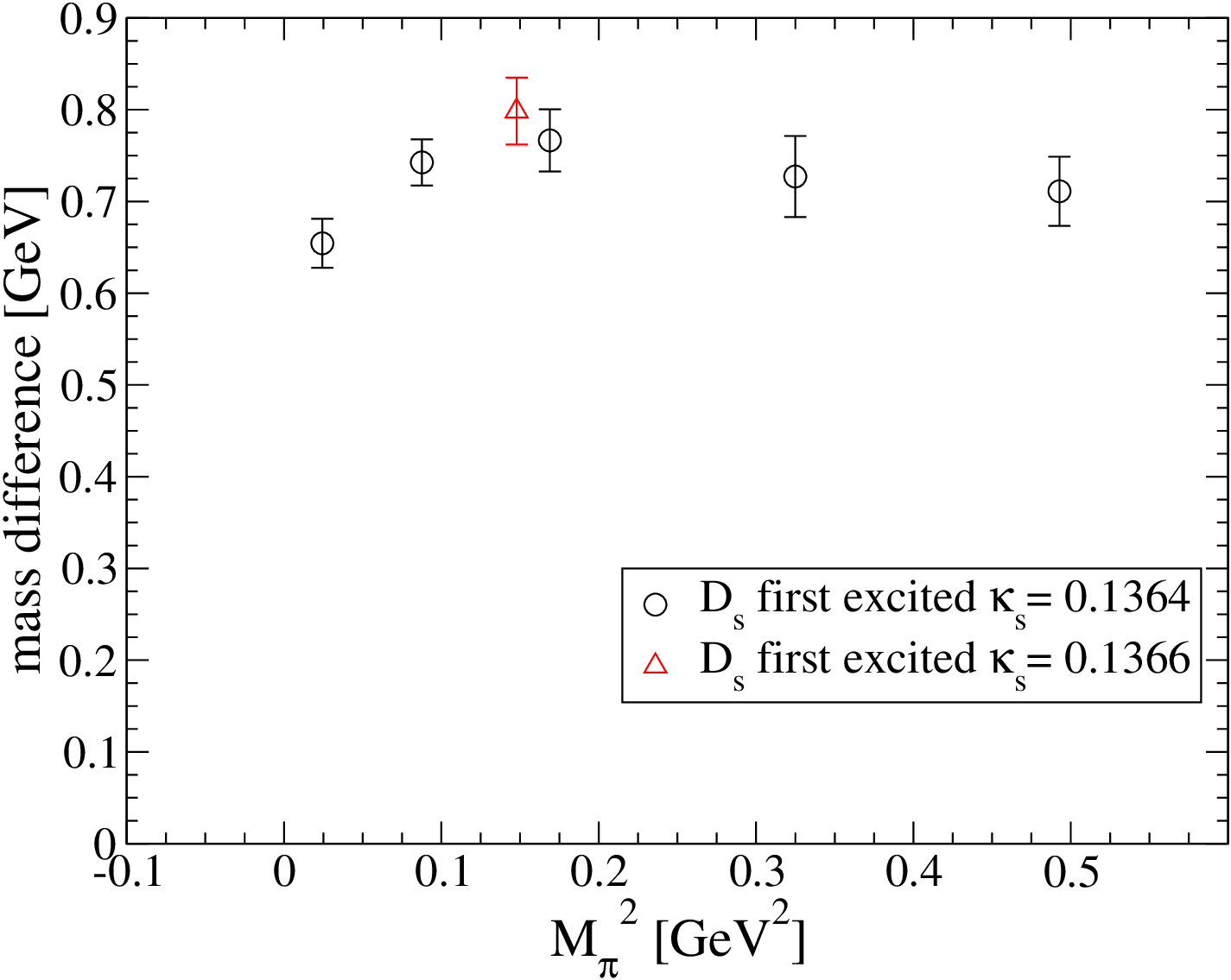}\\
\includegraphics[width=85mm,clip]{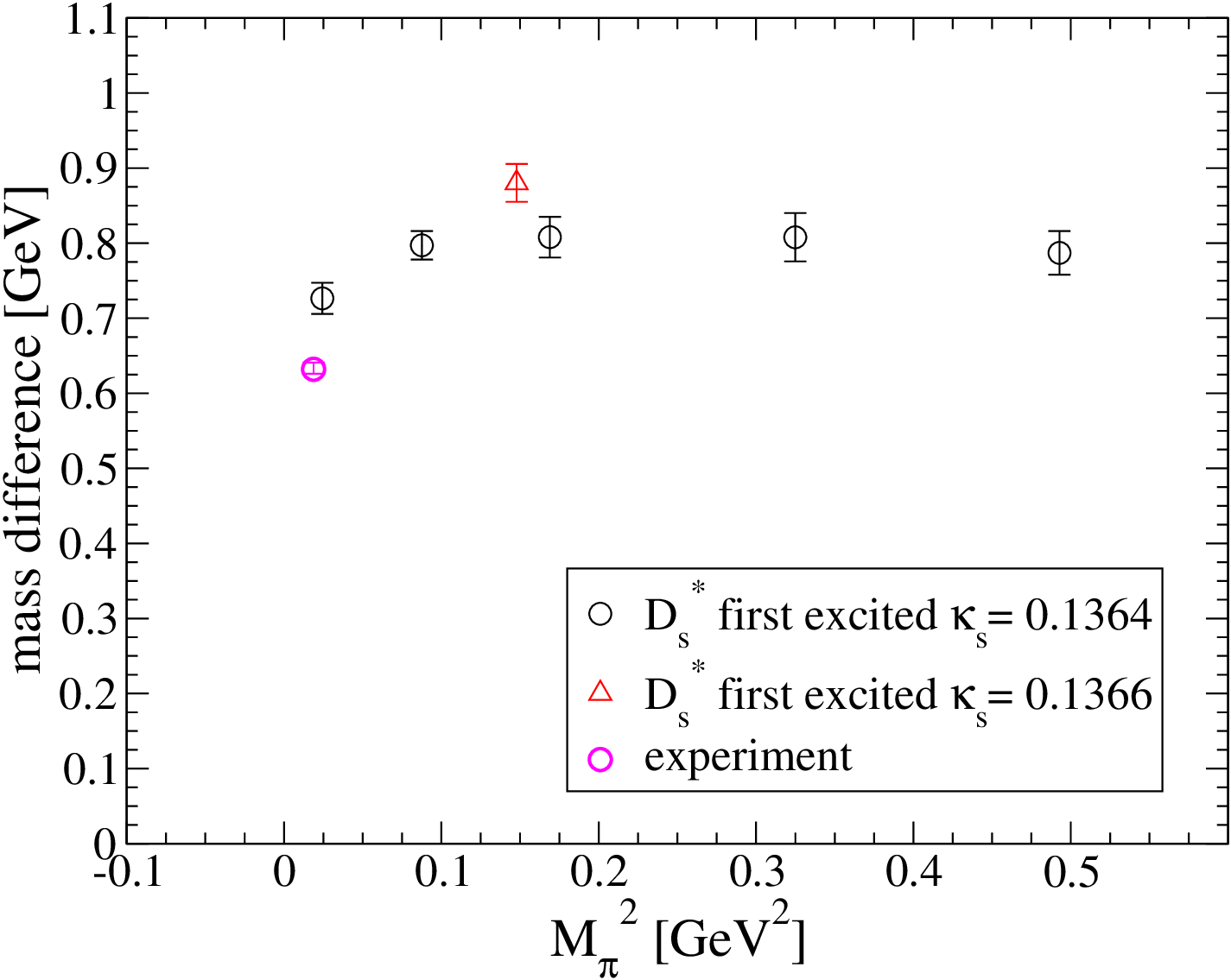}
\caption{Splitting between the 1S and 2S $D_s$ mesons. The top panel shows the splitting between the spin-averaged $\overline{1S}$ and $\overline{2S}$ states. The remaining panels show the individual splittings between the excited $D_s$ (middle panel) and $D_s^\star$ (bottom panel) and the 1S spin-average.  Only one of these states has been clearly identified by experiment. The error displayed is statistical only. The uncertainty in the lattice scale implies that all points could be moved uniformly by $\approx$1.4\%.} 
\label{hl_s1_s2}
\end{figure}

\begin{figure}[t]
\includegraphics[width=85mm,clip]{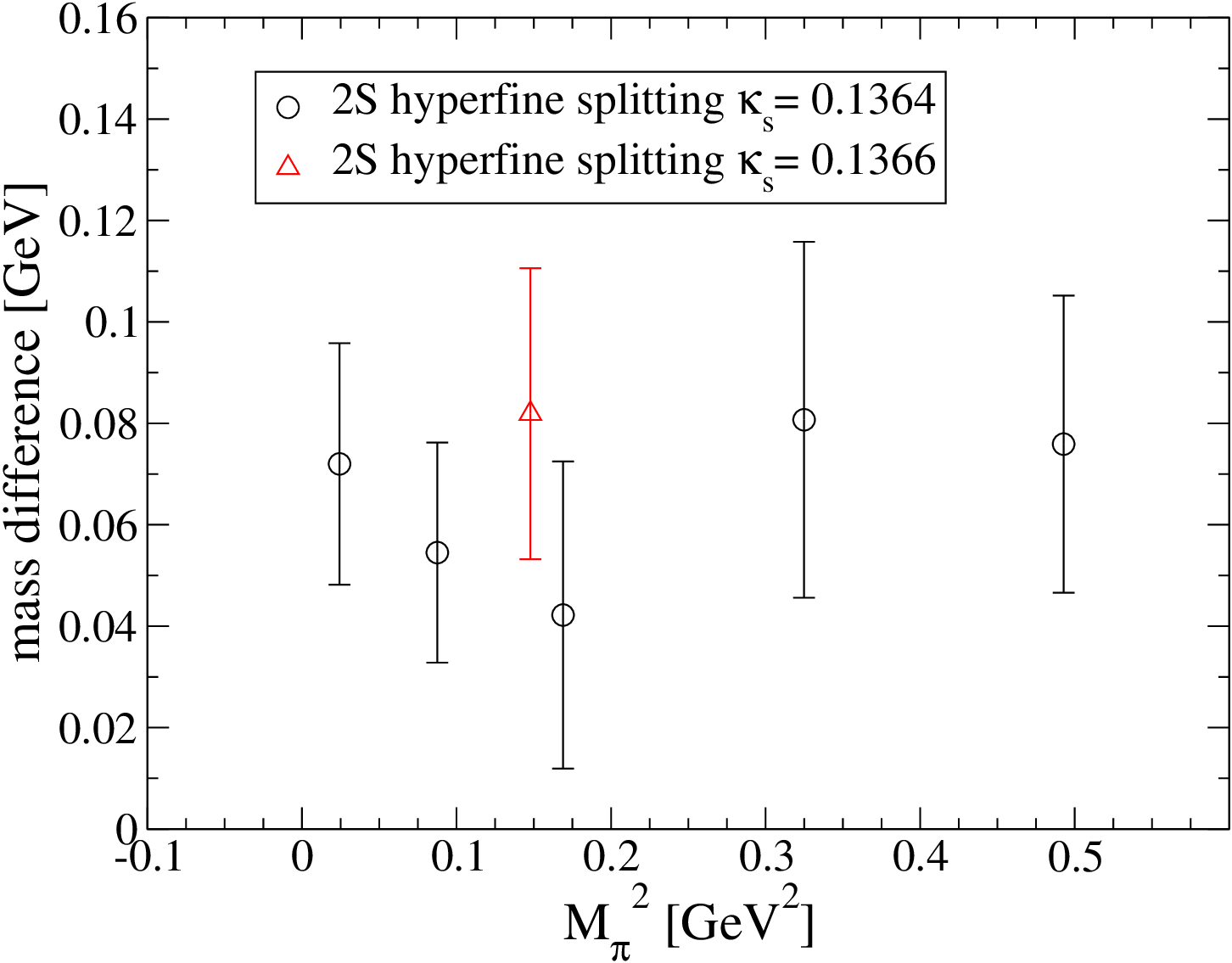}
\caption{Hyperfine-splitting between the $D_s$ 2S states. Errors are statistical only.} 
\label{hs_2s_hyperfine}
\end{figure}

For the $D_s$ mesons, it was possible to extract some information about excited states in the $0^-$ and $1^-$ channels. The top panel of Figure \ref{hl_s1_s2} shows the resulting splitting between the spin-averaged 1S and 2S states. To compare with experiment we also plot the single excited $D_s$ and $D_s^\star$ states in the middle and bottom panels of Figure \ref{hl_s1_s2}. The displayed errors are statistical only. For both 2S states, the result for the splitting at the lightest quark mass is substantially smaller than on all other ensembles. On ensemble 4 with $\kappa_s=0.1366$ the results turn out to be somewhat larger than for all other ensembles. From experiment, a $1^-$ excited state, the $D_{s1}^\star(2700)$ is known. The result at the lowest quark mass is somewhat larger than experiment, but seems reasonable given the statistical and systematic uncertainties, which are larger for excited states than for the ground states. (Regarding the importance of possible multihadron thresholds, remarks similar to those about the $D_{s0}^\star$ and $D_{s1}$ ground states apply.) To complete the analysis of the 2S states, we also plot the 2S hyperfine-splitting in Figure \ref{hs_2s_hyperfine}. In general one expects that the hyperfine-splitting decreases for states higher up in the spectrum. In quark models, (see for example \cite{Godfrey:1985xj}) a typical splitting for the 2S states would be $\approx 60\mathrm{MeV}$. For $D$ mesons the BaBar collaboration recently identified candidates \cite{delAmoSanchez:2010vq} for the corresponding states and the splitting between those $D$ meson states is of comparable size.  While the errors are large compared to the small splitting, we obtain values consistent with these expectations.

\section{\label{conclusion}Summary \& discussion} 

\begin{figure}[tbh]
\includegraphics[width=85mm,clip]{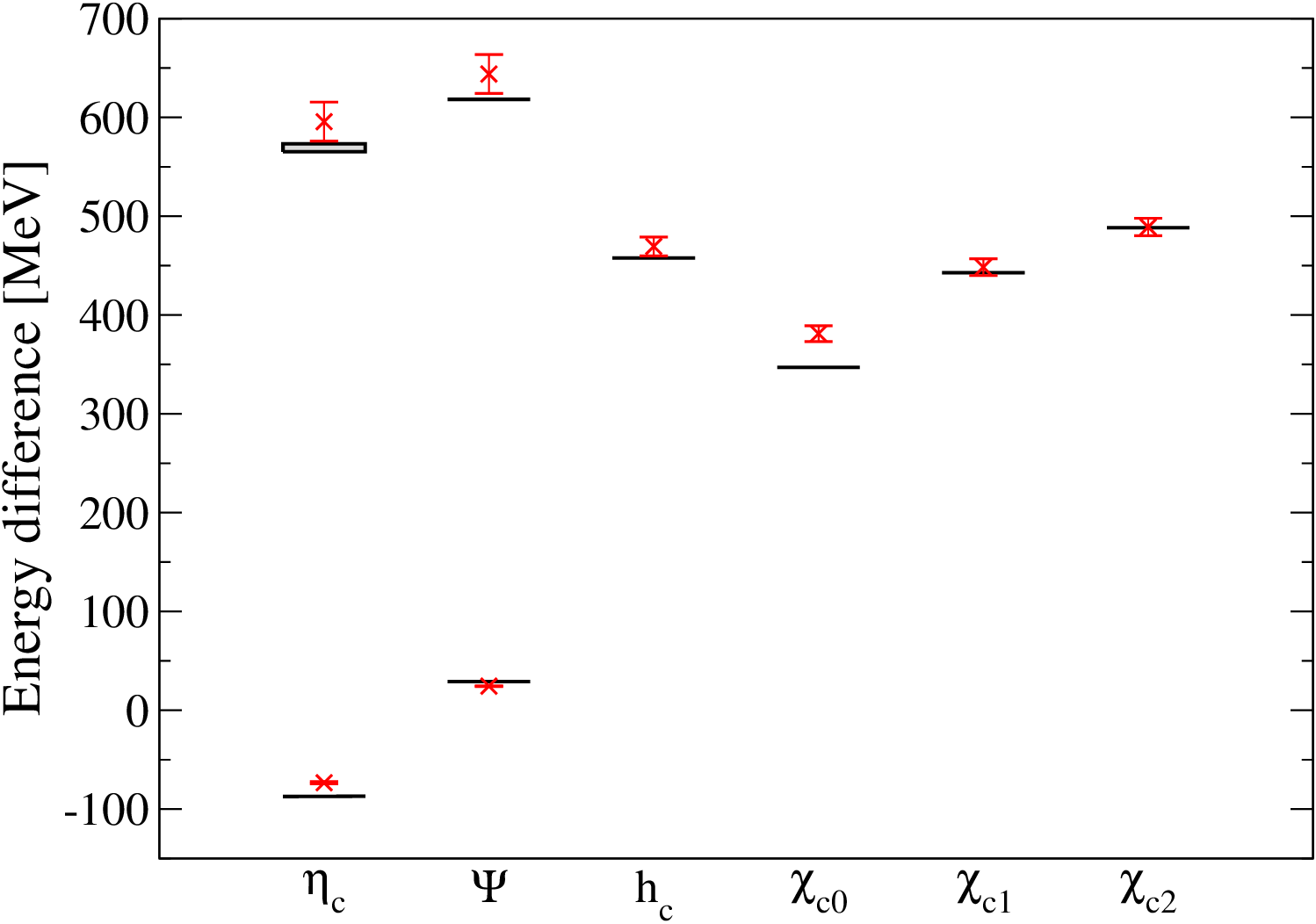}
\caption{Mass differences in MeV for single charmonium states compared to experimental values. All splittings are with respect to the spin-averaged ground state mass $(M_{\eta_c}+3M_{J\Psi})/4$. The errors in the plot have been calculated by combining the errors listed in Table \ref{charmonium_table} in quadrature.} 
\label{charmonium_results}
\end{figure}

We presented results from a dynamical lattice QCD calculation of heavy-light mesons. To investigate the effects of light dynamical quarks, configurations generated by the PACS-CS collaboration \cite{Aoki:2008sm}, with light quarks corresponding to a pion mass as light as 156 MeV, have been used. To test the computational setup, we performed a calculation of the low-lying charmonium spectrum, where scattering thresholds are less important and previous calculations within the same framework exist \cite{Bernard:2010fr}. The results from the ensemble with a pion mass close to the physical pion are summarized in Figure \ref{charmonium_results}. The qualitative features of the low-lying charmonium spectrum are well reproduced which gives us confidence in the procedure used to put the heavy charm quark on the lattice. The numerical values for charmonium mass differences and estimates of some of the uncertainties are provided in Table \ref{charmonium_table}. With a calculation at only one lattice spacing it is not possible to make a quantitative estimate of discretization effects.  However, the expected truncation errors for the Fermilab action as presented in Figs. 3 and 4 of \cite{Oktay:2008ex} can provide a rough guide. With the action used here, discretization effects in spin-averaged mass differences of a few percent could be expected. For more sensitive quantities, e.g., the hyperfine-splitting, only simulation can provide a reliable estimate.

\begin{table}[bht]
\begin{center}
\begin{ruledtabular}
\begin{tabular}{|c|c|c|}
 \T\B Mass difference & This paper [MeV] & Experiment [MeV]\\
\hline
\hline
$\chi_{c0}-\overline{1S}$ & $381.1\pm5.8\pm5.5$ & $347.0\pm0.4$\\
\hline
$\chi_{c1}-\overline{1S}$ & $448.4\pm5.6\pm6.4$ & $442.9\pm0.3$\\
\hline
$\chi_{c2}-\overline{1S}$ & $489.1\pm5.4\pm7.0$ & $488.4\pm0.3$\\
\hline
$h_c-\overline{1S}$ & $469.4\pm6.9\pm6.7$ & $457.7\pm0.4$\\
\hline
$\overline{1P}-\overline{1S}$ & $463.5\pm5.0\pm6.6$ &$457.5\pm0.3$\\
\hline
$\eta_c^\prime-\overline{1S}$ & $595.7\pm17.8\pm8.5$ & $569.2\pm4.0$\\
\hline
$J/\Psi^\prime-\overline{1S}$ & $643.9\pm17.4\pm9.2$ & $618.3\pm0.3$\\
\hline
$\overline{2S}-\overline{1S}$ & $631.8\pm15.6\pm9.0$ & $606.1\pm1.0$\\
\hline
1S hyperfine & $97.8\pm0.5\pm1.4$ & $116.6\pm1.2$\\
\hline
1P spin-orbit & $37.5\pm2.4\pm0.5$ & $46.6\pm0.1$\\
\hline
1P tensor & $10.44\pm1.13\pm0.15$ & $16.25\pm0.07$\\
\hline
2S hyperfine & $48\pm18\pm1$& $49\pm4$\\
\end{tabular}
\end{ruledtabular}
\end{center}
\caption{\label{charmonium_table}Mass differences in the charmonium spectrum in MeV compared to experimental values (calculated from \cite{0954-3899-37-7A-075021}). Bars denote spin-averaged values as discussed in Section \ref{charmonium}. For the results of this paper, the first error denotes the statistical error and the second error denotes the error from setting the lattice scale. In addition there is a non-negligible error from the uncertainty in the determination of $\kappa_c$ for all spin-dependent quantities. We estimated this error for the hyperfine-splitting in the heavy-strange system, but do not determine this uncertainty for charmonium. We stress again that the gauge ensembles at our disposal do not allow for a continuum and infinite volume extrapolation. Consequently we expect qualitative but not quantitative agreement.}
\end{table}

\begin{figure}[tbh]
\includegraphics[width=85mm,clip]{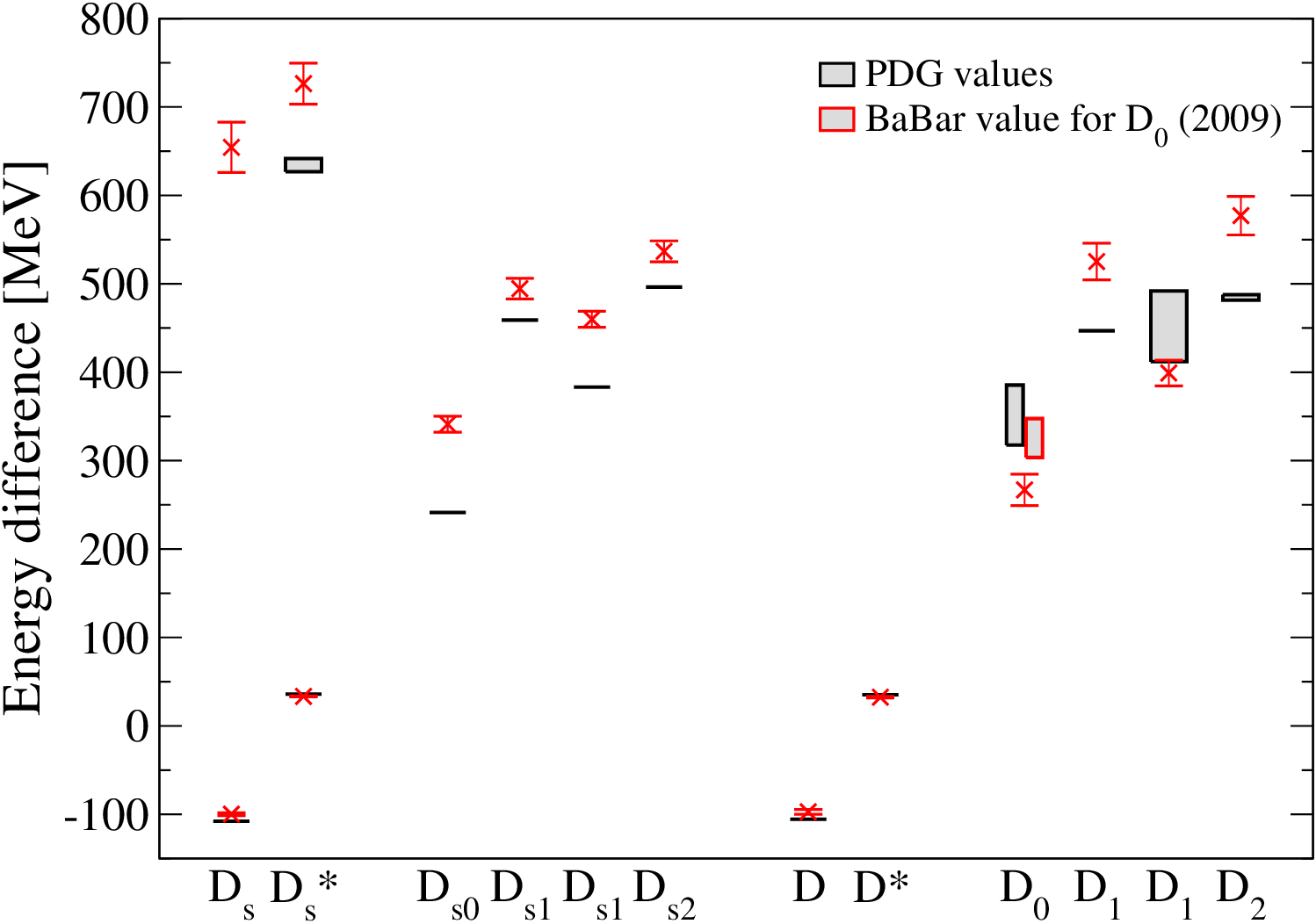}
\caption{Mass differences in MeV for $D$ and $D_s$ meson states compared to experimental values. The difference is always between the state and the spin-averaged ground states $\overline{1S}$ with the same quark content. The errors in the plot have been calculated by combining the errors listed in Table \ref{D_Ds_table} in quadrature. For the $D$ mesons the PDG numbers for the neutral states, which are more accurately determined, are plotted.} 
\label{D_Ds_results}
\end{figure}

Figure \ref{D_Ds_results} shows the results for heavy-light $D$ and heavy-strange $D_s$ mesons as calculated on the ensemble with the lightest quark masses. In both cases differences between the state of interest and the spin-averaged ground state are shown. Table \ref{D_Ds_table} provides numerical values for these splittings and for the 1S and 2S (for the $D_s$) hyperfine-splittings. Even at pion masses close to the physical pion mass, significant differences in comparison to experiment remain for the case of $D_s$ mesons. This is especially evident for the states corresponding to a $j^P=\frac{1}{2}^+$ doublet in the heavy quark limit. For the $D$ mesons, errors are larger and the difference from experimental values is more pronounced for the $j^P=\frac{3}{2}^+$ states. The results for the  $j^P=\frac{1}{2}^+$ doublet on the other hand are qualitatively different from the heavy-strange case.

\begin{table}[bht]
\begin{center}
\begin{ruledtabular}
\begin{tabular}{|c|c|c|}
 \T\B Mass difference & This paper [MeV] & Experiment [MeV]\\
\hline
\hline
$D_{s0}(2317)-\overline{1S}$ & $341.2\pm7.7\pm4.8$ & $241.5\pm0.8$\\
\hline
$D_{s1}(2460)-\overline{1S}$ & $459.8\pm6.4\pm6.4$ & $383.2\pm0.8$\\
\hline
$D_{s1}(2536)-\overline{1S}$ & $494.6\pm9.2\pm6.9$ & $459.0\pm0.5$\\
\hline
$D_{s2}(2573)-\overline{1S}$ & $536.7\pm9.2\pm7.5$ & $496.3\pm1.0$\\
\hline
$D^\prime-\overline{1S}$ & $654.4\pm26.7\pm9.2$ & - \\
\hline
$D^{\star\,\prime}-\overline{1S}$ & $726.4\pm20.8\pm10.2$ & $632.7(+9)(-6)$\\
\hline
$\overline{2S}-\overline{1S}\, D_s$ & $708.4\pm19.9\pm9.9$ & - \\
\hline
1S hyperfine $D_s$ & $133.1\pm1.0\pm1.9$ & $143.8\pm0.4$\\
\hline
2S hyperfine $D_s$ & $72\pm24\pm1$& - \\
\hline
\hline
$D_{0}(2400)-\overline{1S}$ & $266.9\pm17.3\pm3.7$ & $347\pm29$\\
\hline
$D_{1}(2420)-\overline{1S}$ & $399.1\pm13.5\pm5.6$ & $451.6\pm0.6$\\
\hline
$D_{1}(2430)-\overline{1S}$ & $525.2\pm19.4\pm7.4$ & $456\pm40$\\
\hline
$D_{2}(2460)-\overline{1S}$ & $577.1\pm20.3\pm8.1$ & $491.4\pm1.0$\\
\hline
1S hyperfine $D$ & $130.8\pm3.2\pm1.8$ & $140.65\pm0.1$\\
\end{tabular}
\end{ruledtabular}
\end{center}
\caption{\label{D_Ds_table}Mass differences in the $D$ and $D_s$ spectrum in MeV compared to experimental values (calculated from \cite{0954-3899-37-7A-075021}). Bars denote spin-averaged values as discussed in Section \ref{hl_results}. For the results of this paper, the first error denotes the statistical error and the second error denotes the error from setting the lattice scale. In addition there is a non-negligible error from the uncertainty in the determination of $\kappa_c$ for all spin-dependent quantities. For the hyperfine-splitting in the heavy-strange system, this uncertainty is estimated from the tuning run to be 3.2MeV. For a better comparison with the heavy-strange mesons, the hyperfine-splitting for the charged $D$ mesons is used. We stress again that the gauge ensembles at our disposal do not allow for a continuum and infinite volume extrapolation.}
\end{table}

While light dynamical quarks improve the agreement with experiment for heavy-strange and heavy-light mesons, considerable discrepancies remain in the case of the $0^+$ and $1^+$ channels. There are a number of possible reasons for this. For one, future investigations will have to include a full continuum extrapolation. At our lattice spacing, discretization errors are non-negligible. However, in the light of previous results for the charmonium spectrum, it seems unlikely that discretization errors alone are to blame for the observed behavior. Another possible reason for discrepancies is the lack of an extrapolation to the infinite volume limit. While the effects of finite volume are small at large sea quark masses, there might be a concern that this is not the case for the ensemble with the lightest sea quark where $M_\pi L \approx 2.9fm.$ However, finite volume effects have been studied for heavy mesons in \cite{Colangelo:2010ba} and for the ground state D meson are found to be smaller than 1\% for the pion mass and lattice volume used here. Overall the following picture emerges: Given the limitations of the current analysis  the $j^p=\frac{1}{2}^-$ and $j^P=\frac{3}{2}^+$ multiplets are in reasonable agreement with experiment, while the large masses of the heavy-strange $j^P=\frac{1}{2}^+$ doublet resulting from our calculation are hard to explain by these limitations. In particular, we consider it unlikely that a continuum extrapolation will change these results qualitatively.

\begin{figure}[tbh]
\includegraphics[width=85mm,clip]{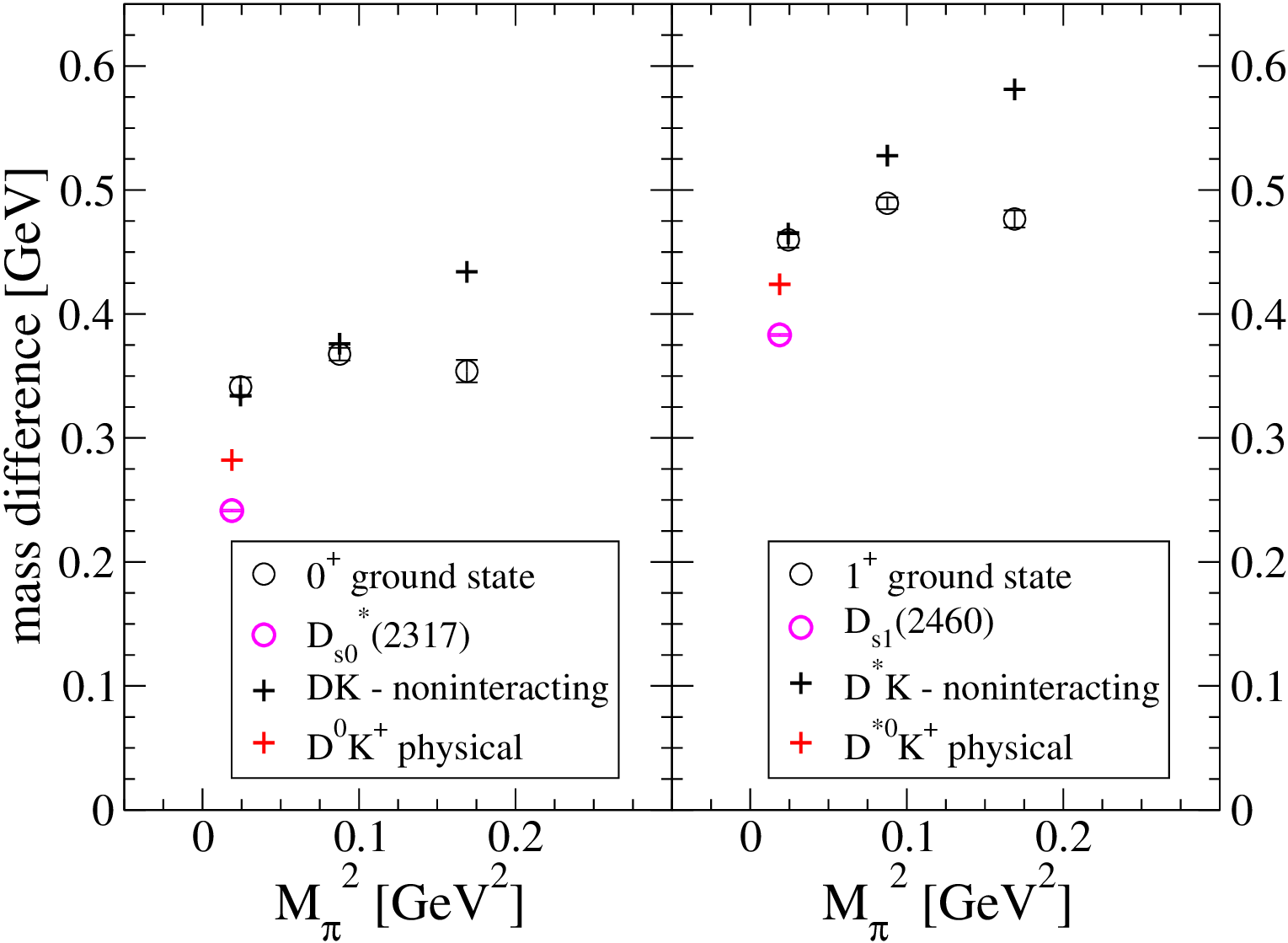}
\caption{Measured energy levels for the $D_{s0}^\star$ (left panel) and $D_{s1}$ (right panel) ground states (black circles) compared to experimental states (magenta circles). All masses are plotted with respect to the spin-averaged $D_s$ ground state. The plus signs denote the $DK$ and $D^\star K$ scattering levels on the lattice (black) and in nature (red). At our lowest pion mass the artificially heavy scattering states are very close to the measured ground state energy.} 
\label{hs_scattering}
\end{figure}

In the case of the heavy-strange $j^p=\frac{1}{2}^-$ states, the $DK$ and $D^\star K$ thresholds may play an important role. Figure \ref{hs_scattering} shows the lowest energy levels for the $D_{s0}^\star$ (left panel) and $D_{s1}$ (right panel) ground states. The physical $DK$ and $D^\star K$ thresholds are indicated with a red plus. The black plus signs show the central value of these scattering states on the ensembles with the three lightest sea quark masses. While the $D$ mesons have been determined in this paper, the kaon mass has been taken from \cite{Aoki:2008sm}. The leading discrepancy between the scattering level at the physical point and the scattering level on the lattice results from the unphysically heavy strange quark mass used in our simulations. An important observation is that the overlap of regular quark-antiquark interpolating fields on multihadron states (which has to be present in dynamical simulations) is most likely small and current state of the art spectrum calculations \cite{Dudek:2010wm,Engel:2010my} do not see most of the possible scattering states with a simple $\bar{q}q$ basis. (For states in a P-wave, the contribution from scattering states is expected to be volume suppressed. However, for many channels even the expected S-wave levels are absent.) Comparing the eigenvectors at our lightest and third-lightest pion mass suggests that we observe the same state on all ensembles and that the expected S-wave scattering level is also absent from our data. The obvious solution is to include multihadron states in the basis for the variational method, which is a challenging task left for future calculations. Including multihadron states in the basis would also enable one to treat these mesons correctly as resonances. The mass and width of hadronic resonances below the inelastic threshold can be determined within L\"uscher's finite volume framework \cite{Luscher:1986pf,Luscher:1990ux}. First attempts to include scattering states for the calculation of the $\rho$ resonance mass and width have been published in \cite{Aoki:2007rd,Feng:2010es,Aoki:2010hn,Frison:2010ws} and similar studies are currently underway for other light-quark resonances. For charmonium excitations, multiparticle states have been included in \cite{Bali:2009er}. For heavy-light mesons first steps in this direction have been made in \cite{Liu:2008rza}.

Finally, there is also the possibility that the states in question are indeed not conventional quark-antiquark states, but rather tetraquark states or other states with which our basis of interpolating fields has poor overlap. For the case of the $D_{s0}^\star$ as a tetraquark, this has been investigated in lattice QCD recently \cite{Gong:2011nr} and no low-lying tetraquark states have been identified.

\acknowledgments
We thank the PACS-CS collaboration for access to their gauge configurations and Martin L\"uscher for making his DD-HMC software available. The calculations were performed on computing clusters at TRIUMF and York University. We thank Sonia Bacca and Randy Lewis for making these resources available. D.M. would like to thank Carleton DeTar, Georg Engel and Sasa Prelovsek for helpful discussions. This work is supported in part by the Natural Sciences and Engineering Research Council of Canada (NSERC).

\begin{appendix}
\section{Tables}
\subsection{\label{source_tables}Interpolating fields}
Table \ref{interpolator_table} collects the interpolators used for both charmonium and heavy-light mesons. For the details of the source construction, please refer to Section \ref{sources}.

\begin{table*}[bht]
\begin{center}
\begin{ruledtabular}
\begin{tabular}{|c|c|c|c|}
  Lattice irrep & Quantum numbers $J^P$ in irrep & Interpolator label & Operator \\
\hline
 $A1^-$ & $0^-$, $4^-$, $\dots$ & A & $\bar{q}_s\gamma_5q^\prime_s$ \\
  & & B & $\bar{q}_s\gamma_t\gamma_5q^\prime_s$ \\
  & & C & $\bar{q}_s\gamma_t\gamma_i\gamma_5q^\prime_{D_i}$ \\
  & & D & $\bar{q}_s\gamma_i\gamma_5q^\prime_{D_i}$ \\
\hline
 $A1^+$ & $0^+$, $4^+$, $\dots$ & A & $\bar{q}_sq^\prime_s$ \\
  & & B & $\bar{q}_s\gamma_iq^\prime_{D_i}$ \\
  & & C & $\bar{q}_s\gamma_t\gamma_iq^\prime_{D_i}$ \\
\hline
 $T_1^-$ & $1^-$, $3^-$, $4^-$, $\dots$ & A & $\bar{q}_s\gamma_iq^\prime_s$ \\
 & & B &  $\bar{q}_s\gamma_t\gamma_iq^\prime_s$ \\
 & & C & $\bar{q}_sq^\prime_{D_i}$\\
 & & D &  $\bar{q}_s\epsilon_{ijk}\gamma_j\gamma_5q^\prime_{D_k}$\\
 & & E & $\bar{q}_s\gamma_tq^\prime_{D_i}$\\
 & & F & $\bar{q}_s\epsilon_{ijk}\gamma_t\gamma_j\gamma_5q^\prime_{D_k}$\\
\hline
 $T_1^+$ & $1^+$, $3^+$, $4^+$, $\dots$ & A & $\bar{q}_s\gamma_i\gamma_5q^\prime_s$ \\
 & & B & $\bar{q}_s\epsilon_{ijk}\gamma_jq^\prime_{D_k}$\\
 & & C & $\bar{q}_s\epsilon_{ijk}\gamma_t\gamma_jq^\prime_{D_k}$\\
 & & D & $\bar{q}_s\gamma_t\gamma_i\gamma_5q^\prime_s$\\
 & & E & $\bar{q}_s\gamma_5q^\prime_{D_i}$\\
 & & F & $\bar{q}_s\gamma_t\gamma_5q^\prime_{D_i}$\\
\hline
 $T_2^-$ & $2^-$, $3^-$, $4^-$, $\dots$ & A & $\bar{q}_s|\epsilon_{ijk}|\gamma_j\gamma_5q^\prime_{D_k}$\\
 & & B & $\bar{q}_s|\epsilon_{ijk}|\gamma_t\gamma_j\gamma_5q^\prime_{D_k}$\\
\hline
 $T_2^+$ & $2^+$, $3^+$, $4^+$, $\dots$ & A & $\bar{q}_s|\epsilon_{ijk}|\gamma_jq^\prime_{D_k}$\\
 & & B & $\bar{q}_s|\epsilon_{ijk}|\gamma_t\gamma_jq^\prime_{D_k}$\\
\hline
 $E^-$ & $2^-$, $4^-$, $\dots$ & A & $\bar{q}_sQ_{ijk}\gamma_j\gamma_5q^\prime_{D_k}$\\
 & & B & $\bar{q}_sQ_{ijk}\gamma_t\gamma_j\gamma_5q^\prime_{D_k}$\\
\hline
 $E^+$ & $2^+$, $4^+$, $\dots$ & A & $\bar{q}_sQ_{ijk}\gamma_jq^\prime_{D_k}$\\
 & & B & $\bar{q}_sQ_{ijk}\gamma_t\gamma_jq^\prime_{D_k}$\\
\end{tabular}
\end{ruledtabular}
\end{center}
\caption{\label{interpolator_table}Table of interpolating field operators sorted by irreducible representation of the lattice symmetry group. The reduced lattice symmetry implies an infinite number of continuum spins in each irreducible representation of the octahedral group. For operators, repeated roman indices indicate summation. The quantity $\gamma_t$ denotes the Dirac matrix for the time direction. The quarks, which are in general not of the same flavor, are referred to by $q$ and $q^\prime$. The subscripts $s$ and $D_i$ denote Gaussian smeared and derivative quark sources respectively. For charmonium, only a subset of operators has been used. Please refer to Section \ref{charmonium} for some cautionary remarks regarding Charmonium and charge conjugation.}
\end{table*}

\subsection{\label{tables_b}Results}

In this appendix, the combination of interpolating fields used for the fits, the respective fit intervals $[t_{min},t_{max}]$ and results for the energy levels are tabulated in Tables \ref{charmonium_fits}, \ref{Ds_fits} and \ref{D_fits}. The ensembles are labeled by their number in Table \ref{paratable}. We also provide the resulting $\chi^2/\mathrm{d.o.f.}$ of the correlated fits and the reference time slice $t_0$ whenever a matrix of interpolators was used. The states are labeled by their quantum numbers. Excited states are listed right below the respective ground states. For a full list of interpolating fields please refer to Section \ref{source_tables}.

\begin{table*}[bhtp]
\begin{center}
\begin{ruledtabular}
\begin{tabular}{|c|c|c|c|c|c|c|c|}
  State &  Interpolators & Ensemble & $t_{min}$ & $t_{max}$ & Energy & $\chi^2/\mathrm{d.o.f.}$ & $t_0$ \\
\hline
 $0^{-+}$ ground state & A,B,C & 1 & 7 & 26 & 1.22924(23) & 1.22 & 4 \\
 &  & 2 & 7 & 27 & 1.22098(21) & 2.03 & 4 \\
 &  & 3 & 7 & 27 & 1.21369(20) & 0.75 & 4 \\
 &  & 4 & 7 & 27 & 1.21017(24) & 0.71 & 4 \\
 &  & 5 & 7 & 27 & 1.21268(15) & 0.71 & 4 \\
 &  & 6 & 7 & 27 & 1.20765(21) & 1.82 & 4 \\
\hline
 $0^{-+}$ first excitation & A,B,C & 1 & 7 & 11 & 1.5595(142) & 1.21 & 4 \\
 &  & 2 & 7 & 15 & 1.5341(136) & 0.53 & 4 \\
 &  & 3 & 7 & 14 & 1.5395(97) & 0.48 & 4 \\
 &  & 4 & 7 & 12 & 1.5247(122) & 2.12 & 4 \\
 &  & 5 & 6 & 14 & 1.5385(71) & 1.78 & 4 \\
 &  & 6 & 6 & 13 & 1.5151(82) & 0.59 & 4 \\
\hline
 $1^{--}$ ground state & A,B,C,D & 1 & 7 & 26 & 1.27877(37) & 1.01 & 4 \\
 &  & 2 & 7 & 27 & 1.26860(33) & 2.08 & 4 \\
 &  & 3 & 7 & 27 & 1.25988(31) & 1.39 & 4 \\
 &  & 4 & 7 & 27 & 1.25499(34) & 0.99 & 4 \\
 &  & 5 & 7 & 27 & 1.25904(22) & 0.93 & 4 \\
 &  & 6 & 7 & 27 & 1.25258(35) & 1.45 & 4 \\
\hline
 $1^{--}$ first excitation & A,B,C,D & 1 & 7 & 15 & 1.5876(110) & 0.85 & 4 \\
 &  & 2 & 7 & 15 & 1.5824(110) & 0.83 & 4 \\
 &  & 3 & 7 & 13 & 1.5535(95) & 0.09 & 4 \\
 &  & 4 & 7 & 12 & 1.5467(99) & 3.16 & 4 \\
 &  & 5 & 7 & 13 & 1.5618(71) & 2.35 & 4 \\
 &  & 6 & 7 & 13 & 1.5373(86) & 0.83 & 4 \\
\hline
 $0^{++}$ ground state & A,B,C & 1 & 7 & 24 & 1.4597(29) & 0.46 & 3 \\
 &  & 2 & 7 & 22 & 1.4468(27) & 1.83 & 3 \\
 &  & 3 & 7 & 23 & 1.4256(29) & 0.57 & 3 \\
 &  & 4 & 7 & 23 & 1.4223(29) & 0.62 & 3 \\
 &  & 5 & 7 & 23 & 1.4289(18) & 0.33 & 3 \\
 &  & 6 & 7 & 24 & 1.4165(28) & 0.70 & 3 \\
\hline
 $1^{++}$ ground state & A & 1 & 7 & 24 & 1.4937(30) & 0.73 & - \\
 &  & 2 & 7 & 25 & 1.4756(30) & 1.41 & - \\
 &  & 3 & 7 & 24 & 1.4584(27) & 0.72 & - \\
 &  & 4 & 7 & 25 & 1.4538(26) & 0.75 & - \\
 &  & 5 & 7 & 24 & 1.4639(22) & 0.69 & - \\
 &  & 6 & 7 & 26 & 1.4475(27) & 0.77 & - \\
\hline
 $2^{++}$ ground state & A,B in T2 irrep& 1 & 7 & 22 & 1.5095(34) & 0.97 & 3 \\
 &  & 2 & 7 & 22 & 1.4961(33) & 0.74 & 3 \\
 &  & 3 & 7 & 21 & 1.4808(29) & 0.86 & 3 \\
 &  & 4 & 7 & 26 & 1.4735(27) & 0.96 & 3 \\
 &  & 5 & 7 & 24 & 1.4856(22) & 0.91 & 3 \\
 &  & 6 & 7 & 24 & 1.4661(26) & 0.88 & 3 \\
\hline
 $1^{+-}$ ground state & D & 1 & 7 & 21 & 1.4932(31) & 0.80 & - \\
 &  & 2 & 7 & 22 & 1.4826(39) & 0.98 & - \\
 &  & 3 & 7 & 24 & 1.4670(31) & 0.39 & - \\
 &  & 4 & 7 & 21 & 1.4627(34) & 1.19 & - \\
 &  & 5 & 7 & 24 & 1.4711(27) & 0.97 & - \\
 &  & 6 & 7 & 25 & 1.4571(33) & 1.38 & - \\
\end{tabular}
\end{ruledtabular}
\caption{\label{charmonium_fits}Tabulated fit results for charmonium. The states are labeled by their quantum numbers $J^{PC}$. The interpolator labels and associated structures can be found in Appendix \ref{source_tables}. The ensembles are named as in Table \ref{paratable}. The energy levels and the associated $\chi^2/\mathrm{d.o.f.}$ are from fully correlated two parameter fits to the eigenvalues of the generalized eigenvalue problem (GEVP) or to single diagonal correlators. Where applicable, the reference time $t_0$ for the GEVP is also tabulated.}
\end{center}
\end{table*}

\begin{table*}[bhtp]
\begin{center}
\begin{ruledtabular}
\begin{tabular}{|c|c|c|c|c|c|c|c|}
  State &  Interpolators & Ensemble & $t_{min}$ & $t_{max}$ & Energy & $\chi^2/\mathrm{d.o.f.}$ & $t_0$ \\
\hline
 $0^{-}$ ground state & A,B,C & 1 & 6 & 26 & 0.86273(33) & 1.42 & 3 \\
 &  & 2 & 6 & 26 & 0.85158(37) & 1.04 & 3 \\
 &  & 3 & 6 & 24 & 0.84000(36) & 1.16 & 3 \\
 &  & 4 & 7 & 25 & 0.82848(40) & 1.37 & 3 \\
 &  & 5 & 6 & 24 & 0.83929(26) & 1.22 & 3 \\
 &  & 6 & 7 & 25 & 0.83149(34) & 1.80 & 3 \\
\hline
 $0^{-}$ first excitation & A,B,C & 1 & 6 & 11 & 1.2423(175) & 1.69 & 3 \\
 &  & 2 & 6 & 11 & 1.2361(205) & 1.74 & 3 \\
 &  & 3 & 6 & 11 & 1.2406(157) & 1.19 & 3 \\
 &  & 4 & 6 & 12 & 1.2418(168) & 1.65 & 3 \\
 &  & 5 & 6 & 12 & 1.2294(117) & 1.32 & 3 \\
 &  & 6 & 6 & 13 & 1.1782(123) & 1.24 & 3 \\
\hline
 $1^{-}$ ground state & A,B,C,D & 1 & 6 & 28 & 0.93297(59) & 0.73 & 3 \\
 &  & 2 & 6 & 26 & 0.91864(63) & 1.51 & 3 \\
 &  & 3 & 6 & 25 & 0.90429(60) & 1.57 & 3 \\
 &  & 4 & 7 & 27 & 0.89015(69) & 1.20 & 3 \\
 &  & 5 & 6 & 24 & 0.90429(43) & 0.68 & 3 \\
 &  & 6 & 7 & 30 & 0.89268(59) & 1.88 & 3 \\
\hline
 $1^{-}$ first excitation & A,B,C,D & 1 & 6 & 13 & 1.2772(135) & 0.63 & 3 \\
 &  & 2 & 6 & 10 & 1.2732(150) & 0.91 & 3 \\
 &  & 3 & 6 & 12 & 1.2600(126) & 0.95 & 4 \\
 &  & 4 & 6 & 12 & 1.2794(117) & 2.07 & 3 \\
 &  & 5 & 6 & 12 & 1.2545(88) &  1.03 & 3 \\
 &  & 6 & 6 & 13 & 1.2113(96) &  0.56 & 4 \\
\hline
 $0^{+}$ ground state & A,B,C & 1 & 6 & 21 & 1.1057(35) & 0.98 & 3 \\
 &  & 2 & 6 & 21 & 1.0784(40) & 0.59 & 3 \\
 &  & 3 & 6 & 20 & 1.0509(43) & 0.92 & 3 \\
 &  & 4 & 7 & 13 & 1.0399(44) & 1.28 & 3 \\
 &  & 5 & 6 & 24 & 1.0570(25) & 0.68 & 3 \\
 &  & 6 & 6 & 20 & 1.0342(37) & 1.54 & 3 \\
\hline
 $1^{+}$ ground state & A,C,D,F & 1 & 7 & 13 & 1.1574(35) & 0.12 & 4 \\
 &  & 2 & 7 & 15 & 1.1319(35) & 1.22 & 4 \\
 &  & 3 & 7 & 15 & 1.1073(34) & 1.01 & 4 \\
 &  & 4 & 7 & 13 & 1.0907(33) & 2.63 & 4 \\
 &  & 5 & 7 & 15 & 1.1130(23) & 1.04 & 4 \\
 &  & 6 & 7 & 14 & 1.0888(31) & 0.79 & 4 \\
\hline
 $1^{+}$ first excitation & A,C,D,F & 1 & 7 & 13 & 1.1654(54) & 0.82 & 4 \\
 &  & 2 & 7 & 15 & 1.1458(55) & 1.02 & 4 \\
 &  & 3 & 7 & 15 & 1.1273(49) & 0.27 & 4 \\
 &  & 4 & 7 & 13 & 1.1155(51) & 2.92 & 4 \\
 &  & 5 & 7 & 15 & 1.1320(35) & 1.22 & 4 \\
 &  & 6 & 7 & 14 & 1.1047(44) & 1.01 & 4 \\
\hline
 $2^{+}$ ground state & A,B in T2 irrep & 1 & 7 & 20 & 1.1810(52) & 0.76 & 3 \\
 &  & 2 & 7 & 17 & 1.1662(60) & 1.11 & 3 \\
 &  & 3 & 7 & 18 & 1.1491(52) & 0.72 & 3 \\
 &  & 4 & 7 & 18 & 1.1353(51) & 1.47 & 3 \\
 &  & 5 & 7 & 19 & 1.1530(36) & 1.41 & 3 \\
 &  & 6 & 7 & 20 & 1.1241(44) & 1.29 & 3 \\
\end{tabular}
\end{ruledtabular}
\end{center}
\caption{\label{Ds_fits}Tabulated fit results for the $D_s$ mesons. The states are labeled by their quantum numbers $J^{P}$. The interpolator labels and associated structures can be found in Appendix \ref{source_tables}. The ensembles are named as in Table \ref{paratable}. The energy levels and the associated $\chi^2/\mathrm{d.o.f.}$ are from fully correlated two parameter fits to the eigenvalues of the generalized eigenvalue problem (GEVP) or to single diagonal correlators. Where applicable, the reference time $t_0$ for the GEVP is also tabulated.}
\end{table*}

\begin{table*}[bhtp]
\begin{center}
\begin{ruledtabular}
\begin{tabular}{|c|c|c|c|c|c|c|c|}
  State &  Interpolators & Ensemble & $t_{min}$ & $t_{max}$ & Energy & $\chi^2/\mathrm{d.o.f.}$ & $t_0$ \\
\hline
\hline
 $0^{-}$ ground state & A,B,C & 1 & 6 & 26 & 0.84022(37) & 1.52 & 3 \\
 &  & 3 & 7 & 23 & 0.79580(61) & 0.79 & 3 \\
 &  & 5 & 7 & 26 & 0.78798(82) & 1.35 & 3 \\
 &  & 6 & 7 & 24 & 0.77646(119) & 1.23 & 3 \\
\hline
 $1^{-}$ ground state & A,B,C,D & 1 & 5 & 26 & 0.91237(60) & 0.89 & 3 \\
 &  & 3 & 7 & 26 & 0.86327(99) & 0.92 & 3 \\
 &  & 5 & 7 & 22 & 0.85776(122) & 1.41 & 3 \\
 &  & 6 & 7 & 24 & 0.83656(189) & 1.10 & 3 \\
\hline
 $0^{+}$ ground state & A,B,C & 1 & 6 & 15 & 1.0911(50) & 0.28 & 3 \\
 &  & 3 & 6 & 11 & 0.9956(80) & 0.38 & 3 \\
 &  & 5 & 6 & 17 & 1.0150(66) & 1.04 & 3 \\
 &  & 6 & 6 & 18 & 0.9442(79) & 0.68 & 3 \\
\hline
 $1^{+}$ ground state & A,D & 1 & 7 & 14 & 1.1394(47) & 1.09 & 4 \\
 &  & 3 & 7 & 15 & 1.0631(48) & 0.64 & 4 \\
 &  & 5 & 7 & 12 & 1.0601(58) & 2.07 & 4 \\
 &  & 6 & 7 & 15 & 1.0050(63) & 0.84 & 4 \\
\hline
 $1^{+}$ first excitation & A,D & 1 & 7 & 14 & 1.1538(65) & 1.00 & 4 \\
 &  & 3 & 7 & 15 & 1.0874(69) & 0.77 & 4 \\
 &  & 5 & 7 & 12 & 1.0883(78) & 0.65 & 4 \\
 &  & 6 & 7 & 15 & 1.0629(93) & 0.79 & 4 \\
\hline
 $2^{+}$ ground state & A,B in T2 irrep & 1 & 7 & 18 & 1.1869(60) & 0.85 & 3 \\
 &  & 6 & 7 & 18 & 1.1153(74) & 1.48 & 3 \\
 &  & 3 & 7 & 15 & 1.1166(80) & 1.02 & 3 \\
 &  & 5 & 7 & 18 & 1.0868(95) & 1.10 & 3 \\
\end{tabular}
\end{ruledtabular}
\end{center}
\caption{\label{D_fits}Tabulated fit results for the $D$ mesons. The states are labeled by their quantum numbers $J^{P}$. The interpolator labels and associated structures can be found in Appendix \ref{source_tables}. The ensembles are named as in Table \ref{paratable}. The energy levels and the associated $\chi^2/\mathrm{d.o.f.}$ are from fully correlated two parameter fits to the eigenvalues of the generalized eigenvalue problem (GEVP) or to single diagonal correlators. Where applicable, the reference time $t_0$ for the GEVP is also tabulated.}
\end{table*}

\bibliography{bibtex}
\bibliographystyle{h-physrev4}

\end{appendix}

\end{document}